\documentclass[aps,pre,reprint,superscriptaddress,longbibliography,10pt]{revtex4-2}
\usepackage{amsmath, amssymb, graphicx, color, braket,mathtools}
\usepackage{bbold, bm}
\usepackage{verbatim}

\begin{document}
	
\title{{Gate-tunable supercurrent switching in a three-terminal quantum dot transistor}}
	
\author{David Nov\'ak}
\email{david.novak@matfyz.cuni.cz}
\affiliation{Institute of Physics, Czech Academy of Sciences, Na Slovance 2, CZ-18200 Praha 8, Czech Republic}
\affiliation{Department of Condensed Matter Physics, Faculty of Mathematics and Physics, Charles University, Ke Karlovu 5, CZ-121 16 Praha 2, Czech Republic}

\author{Kacper Wrze\'{s}niewski}
\affiliation{Institute of Physics, Czech Academy of Sciences, Na Slovance 2, CZ-18200 Praha 8, Czech Republic}

\author{Filip Krizek}
\affiliation{Institute of Physics, Czech Academy of Sciences, Na Slovance 2, CZ-18200 Praha 8, Czech Republic}
\affiliation{Faculty of Nuclear Sciences and Physical Engineering, Czech Technical University in Prague, 11519 Prague, Czech Republic}

\author{Wolfgang Belzig}
\affiliation{Fachbereich Physik, Universit{\"a}t Konstanz, D-78457 Konstanz, Germany}
	
\author{Peter Zalom}
\email{zalomp@fzu.cz}
\affiliation{Institute of Physics, Czech Academy of Sciences, Na Slovance 2, CZ-18200 Praha 8, Czech Republic}

\date{\today}
	
\begin{abstract}
We propose and analyze a three-terminal superconducting transistor switch in which a gate-defined quantum dot region interrupts a flux-controlled loop. The output supercurrent is switched off completely when the two loop arms couple to the dot with equal tunneling strengths and half a flux quantum threads the loop. The device then operates as a symmetric SQUID at half flux. The condition follows from a single geometric factor $\chi$, which reduces the three-terminal problem to an equivalent two-terminal Anderson impurity model and makes the switching mechanism independent of the Coulomb interaction, the dot level position, the total hybridization and temperature. These enter only through a prefactor and thus decide how tolerant the device is to imperfect barrier tuning. Numerical renormalization group calculations across the full coupling parameter space show that the resonance is embedded in a broad region of elevated switching ratio at intermediate hybridization $\Gamma \approx \Delta$, while both weaker and stronger coupling reduce it to a thin isolated line. Finite temperature smooths the underlying singlet-doublet transition and lowers the supercurrent amplitude, but affects the open and closed states comparably and leaves the switching ratio largely intact. Within a zero-bandwidth two-level model, a second dot level is shown to open a leakage channel that grows as the two levels approach in energy. Raising the output barrier and the realistic presence of inter-level Coulomb repulsion, however, suppress the leakage.
\end{abstract}
	
	
\maketitle
	
\section{Introduction \label{sec:intro} }
	
Semiconductor field-effect transistors form the foundation of present-day digital electronics, with physical gate lengths in high-volume manufacturing reaching $\approx 20$~nm while maintaining on/off current ratios of $10^5$--$10^7$ between the conducting and blocking states~\cite{Lundstrom-2003, Bohr-2017}. However, both the switching and leakage currents of such devices dissipate power, and this constraint severely limits their applications at cryogenic temperatures. A dilution refrigerator provides only hundreds of microwatts of cooling power at $100$~mK, while a single complementary metal–oxide–semiconductor (CMOS) control channel dissipates milliwatts. Control electronics for solid-state qubits is therefore kept mostly at room temperature and connects to the cold stages through coaxial lines that conduct heat and carry thermal photons downward.

Cryogenic CMOS controllers relocate part of this electronics to the $4$~K stage, where more cooling power is available, but this stage remains hot on the scale of the qubit and the signal lines must still be attenuated on their way down~\cite{Pauka-2021}. Superconducting (SC) circuit elements operating natively at the qubit stage would remove this thermal boundary entirely, making SC classical logic a natural candidate for integration with quantum computation nodes. Such elements operate with supercurrents that must be controllably suppressed (the OFF state) or allowed to flow (the ON state), with a switching ratio of at least several orders of magnitude required for reliable control. While ratios of this size have been reported in gate-controlled suppression of the supercurrent in SC heterostructures~\cite{deSimoni-2018,Paolucci-2018,deSimoni-2019,deSimoni-2020,Puglia-2020,Rocci-2020,Ritter-2021,Rocci-2021}, the microscopic origin of such gate-controlled-supercurrent effects remains under debate~\cite{Bours-2020,Chirolli-transistor-2021,Amoretti-2022, Chakraborty-2023,Zaccone-2024}; see Ref.\cite{Ruf-2024GateControl} for a recent review. SC elements based on Josephson junctions, which switch via the phase rather than the condensate, are thus a more natural pathway to overcome both limitations.

Josephson junctions are often realized either as superconductor--insulator--superconductor (SIS) tunnel junctions, such as Al/AlO$_x$/Al stacks~\cite{Dolan-1977}, or as superconductor--normal-metal--superconductor (SNS) junctions, in which the supercurrent is carried through a weak link~\cite{Likharev-1979}. Electrostatic control of the weak link was first demonstrated in planar junctions defined in GaAs/AlGaAs~\cite{Takayanagi-1985} and InAs~\cite{Nitta-1993} two-dimensional electron gases (2DEGs) with related Josephson field-effect transistor implementations demonstrated subsequently~\cite{Akazaki-1996JosephsonFET}. Later developments involved carbon nanotubes~\cite{Kasumov-1999} and semiconductor nanowires~\cite{Doh-2005}. A complementary approach employs electrostatic gates to control Josephson transport through quantum dots by tuning discrete levels and/or their tunnel coupling to superconducting electrodes~\cite{Herrero-2006Quantum, Cleuziou-2006cnt, Cleuziou-2007GateTuned, Jorgensen-2006FabryPerot}. Among these, planar 2DEGs combine a proximity-induced gap and high mobility with lithographic flexibility; InAs quantum wells with epitaxially grown Al are the most developed realization~\cite{Shabani-2016,Kjaergaard-2016}, with InSb~\cite{Ke-2019}, HgTe~\cite{Bocquillon-2017} and planar Ge~\cite{Vigneau-2019, Luethi-2023,Leblanc-2025} representing emerging alternatives.

The versatility and lithographic flexibility of planar 2DEG platforms extend naturally to multi-terminal geometries~\cite{Cohen-2018,Pankratova-2020,Draelos-2019,Coraiola-2023,Coraiola-2024,Antonelli-2025}, which give access to phenomena with no counterpart in two-terminal devices. These include coherent transport of multiple Cooper pairs~\cite{Jonckheere-2013,Cohen-2018,Freyn-2011,Melin-2023,Zhang-2023} and the engineering of Andreev bound state (ABS) bands in a multi-dimensional phase space~\cite{Riwar-2016,Eriksson-2017,Strambini-2016,Klees-2020,Weisbrich-2021,Coraiola-2023,Coraiola-2024, Coraiola-2024prx}, as well as supercurrent diode and transistor effects. The latter were recently identified as phase-controlled phenomena in three-terminal structures whose central region hosts a single electronic level, modeled as a quantum dot (QD) with non-negligible Coulomb interaction~\cite{vanDam-2006Supercurrent, Jorgensen-2007CriticalCurrent,Zalom-2024}. 

The reported effect was, however, restricted to a single configuration of the QD and its couplings to the SC leads. The resulting switching ratio reached at most one order of magnitude, well short of the requirement set out above. Whether this reflects an intrinsic bound of the multi-terminal mechanism or merely the particular parameters chosen remains open. The device nevertheless offers two physically distinct sets of parameters, both addressable in realistic setups. While the SC phases of the leads, controlled by magnetic fluxes, underlie the transistor effect reported in Ref.~\cite{Zalom-2024}, the electrostatic properties of the QD, together with its tunnel couplings to the SC leads, remain unexplored in this context.


The paper is organized as follows. In Sec.~\ref{sec:multi-trans}, we introduce the three-terminal device architecture in which two SC terminals form a flux-controlled loop interrupted by a QD region, and review in Sec.~\ref{subsec:single_level} the single-level theory of the multi-terminal Anderson model that allows to reduce the complexity of supercurrent calculations by introducing the concept of the geometric factor $\chi$. Section~\ref{sec:sc_trans} contains the central analytic result of the paper: a supercurrent switching mechanism arising when both arms of the loop couple to the QD with equal tunneling strengths. The tolerance of this resonance condition to imperfect barrier tuning is examined in Sec.~\ref{sec:resonance_widths} across the full coupling parameter space, where Sec.~\ref{subsec:origin} traces the observed structures back to the underlying singlet-doublet quantum phase transitions and Sec.~\ref{subsec:temperature} addresses their temperature dependence. Deviations from the ideal single-level scenario are then analyzed within a zero-bandwidth two-level model in Sec.~\ref{sec:zbw}, separately for leakage caused by the energetic proximity of an additional level (Sec.~\ref{subsec:proximity}) and for the role of inter-level Coulomb repulsion (Sec.~\ref{subsec:inter}). In Sec.~\ref{sec:fabrication}, we discuss the technological feasibility of the proposed device within standard lithographic techniques for Al/InAs 2DEG heterostructures. Conclusions are given in Sec.~\ref{sec:conclusions}.

\section{Three-terminal SC transistor device \label{sec:multi-trans}}

\begin{figure}[t]
    \includegraphics[width=\columnwidth]{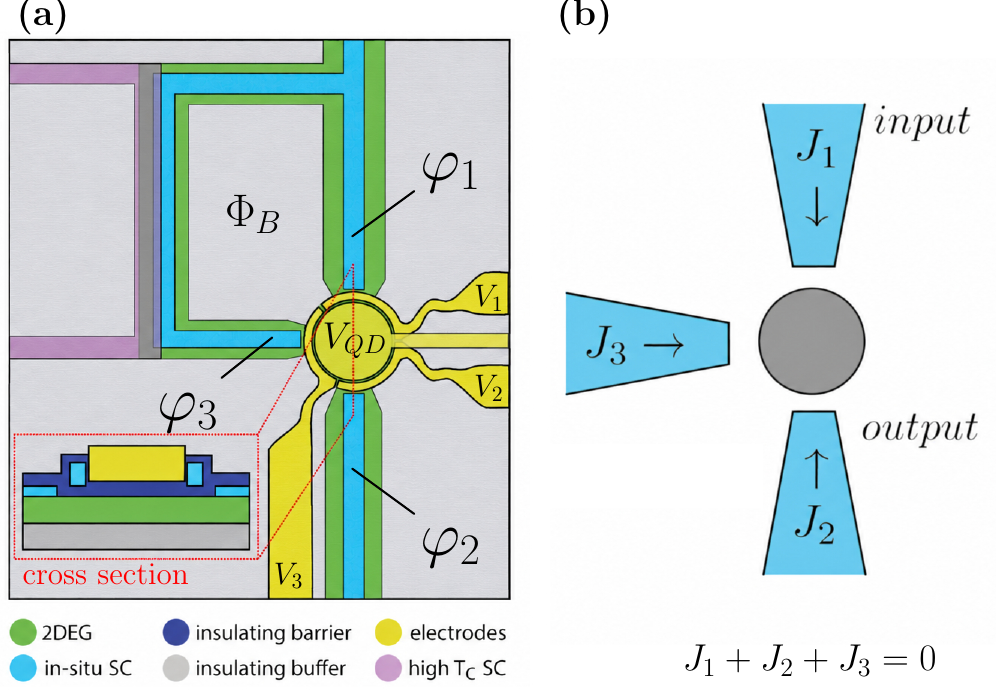}
    \caption{
    (a)
    Scheme of a three-terminal superconducting (SC) quantum-dot (QD) transistor in a planar 2DEG heterostructure. SC terminals $1$ and $3$, with phases $\varphi_1$ and $\varphi_3$, form a flux-controlled loop, while terminal $2$, with phase $\varphi_2$, serves as the output electrode. A magnetic flux $\Phi_{\rm B}$ threading the loop fixes the phase difference $\varphi_3-\varphi_1=2\pi\Phi_{\rm B}/\Phi_0$, where $\Phi_0=h/2e$ is the superconducting flux quantum. The central QD region is defined electrostatically:
    $V_{\rm QD}$ controls the QD potential, whereas $V_1$, $V_2$, and $V_3$ tune the tunnel barriers between the
    QD and the corresponding superconducting terminals. 
    The inset illustrates a cross section of the gate-defined junction, and the color code identifies the 2DEG, in-situ superconductor, insulating barrier and buffer layers, electrostatic gates, and the additional SC layer. Further details on the fabrication requirements are given in Sec.~\ref{sec:fabrication}.
    (b)
    Current convention used throughout the manuscript. The terminal currents $J_j$ are defined as positive when flowing toward the QD region. The incoming current therefore splits between the two arms of the loop as $J_{\rm in}=J_1+J_3$, recombines at the QD, and exits through terminal $2$ as $J_{\rm out}=-J_2$. 
    }
    \label{fig:sc_trans}
\end{figure}

A classical transistor switch comprises input, output, and control (base) terminals, with the latter designated for controlling the input and output currents. While this concept translates readily into the SC scenario as demonstrated in Ref.~\cite{Zalom-2024}, its straightforward application is impractical. In detail, an incoming supercurrent entering the SC transistor through the input electrode would inevitably be split into the remaining SC electrodes. Consequently, the incoming current $J_{\mathrm{in}}$ would not convert fully into the output lead as $J_{\mathrm{out}}=J_{\mathrm{in}}$. Instead, a fraction of $J_{\mathrm{in}}$ would always leak into the base electrode. Crucially, such leakage current would generically be of the same order as the one reaching the output, and would moreover vary with the SC phases. Consequently, the conversion ratio $J_{\mathrm{out}}/J_{\mathrm{in}}$ would neither be fixed nor close to unity. Additionally, in such a conventional design, the OFF state has to be reached through the details of $0$-$\pi$ quantum phase transitions (QPTs), which are device and material
specific~\cite{Zalom-2024}.

We therefore abandon the conventional input-output-base concept and instead connect two SC terminals (originally designated in Ref.~\cite{Zalom-2024} as input and base) to a flux-controlled loop. The motivation for this choice is
the well-known interference effect in a dc SQUID with identical junctions~\cite{Jaklevic-1964, Tesche-1977}. When its loop is threaded by half a flux quantum and the loop inductance is negligible, the supercurrents in the two arms interfere destructively: the transport critical current vanishes, while a current keeps circulating within the loop. A flux-controlled loop therefore provides a natural OFF state that does not rely on material-specific QPTs. In conventional SQUIDs, however, the junctions are fixed by fabrication, and any mismatch between them leaves a residual critical current at half flux~\cite{Fulton-1972, Koch-2007}.

In the present device, the loop is interrupted by a QD region, which acts as the common node of the two arms, as shown in Fig.~\ref{fig:sc_trans}(a). The role of the two SQUID junctions is then taken by the gate-tunable barriers between the QD and terminals $1$ and $3$, so that their symmetry can be restored in situ rather than being fixed by fabrication. A third gate-tunable barrier couples the QD to the output terminal $2$. Since the two loop terminals are joined externally, they present a single current path to the rest of the circuit, and charge conservation alone then guarantees complete channeling of the incoming supercurrent through the device:
\begin{align}
	J_{\mathrm{in}}
	=
	J_{\mathrm{out}}
	=
	-J_2
    =
	J_1+J_3.
	\label{eq:conservation}
\end{align}
Crucially, thanks to the loop architecture the charge conservation ensures that input and output currents are the same irrespective of the dot occupation, the Coulomb interaction, the hybridization strengths, the SC phases, or temperature. The supercurrent passing through the device is then controlled solely via the magnetic flux $\Phi_{\rm{B}}$ penetrating the loop. Assuming a loop inductance small enough that the phase drop generated by the circulating supercurrent is negligible, the flux fixes the phase difference between the two loop terminals as
\begin{align}
	\varphi_3 - \varphi_1
	=
	2\pi \frac{\Phi_{\rm{B}}}{\Phi_0},
	\label{eq:flux}
\end{align}
with $\Phi_0=h/2e$ the SC flux quantum. The flux-controlled loop thus serves a dual purpose: it sets the phase difference entering \eqref{eq:flux} and thereby the operating point of the switch, while simultaneously controlling how the input supercurrent splits between the two arms~\cite{Zalom-2024, Zalom-2026}. In operation, the device is switched between its ON and OFF states by threading integer or half-integer flux quanta through the loop, while the phase $\varphi_2$ across the output branch is set by the external circuit.

In what follows, we label the three SC terminals as $j \in \{1,2,3 \}$. Through each of the terminals, a supercurrent $J_j$ passes in a direction selected so that its positive-values indicate flow towards the QD-lead interface [Fig.~\ref{fig:sc_trans}(b)]. The terminals $1$ and $3$ are designated to form the flux-controlled loop, while the SC terminal $2$ serves as the output electrode. Current conservation requires the input supercurrent $J_{\mathrm{in}}$ to split into supercurrents $J_1$ and $J_3$ upon entering the device. Each of these supercurrents is then passing through the tunnel barrier between the QD region and the corresponding SC terminal. In the QD region, the two supercurrents $J_1$ and $J_3$ recombine into the output supercurrent $J_{\mathrm{out}}$, which due to the selected convention about the flow direction fulfills $J_{\mathrm{out}}=-J_2$. Crucially, its magnitude as well as the split of the incoming supercurrent are governed by quantum phenomena dictated by the multi-terminal Josephson effect.

\subsection{Single-level in the QD region \label{subsec:single_level}}

The transport across the tunneling barriers connecting the SC leads and the QD region is dependent on the properties of the QD region. We first consider a scenario where one tunes the QD region by gate voltage so that a single electronic level resides in the transport window and remaining levels are well separated~\footnote{In 2DEG systems, this level can be selected as the last one to hold electrons before the QD region is fully depleted electrostatically}. We denote the energy of such a dominant level by $\varepsilon_{\rm{d}}$ and consider a repulsive Coulomb interaction $U$ for two electrons occupying it simultaneously. The full Hamiltonian of the three-terminal device then reads
\begin{align}
	H
	&=
	H_{\rm{d}}
	+
	\sum_{j=1}^{3}
	\left(
	H_{j,\rm{SC}}
	+
	H_{j,\rm{T}}
	\right),
    \label{eq:Hsingle}
\end{align}
where $H_{\rm{d}}$ describes the dominant electronic level in the QD region, $H_{j,\rm{SC}}$ is the conventional Bardeen-Cooper-Schrieffer (BCS) Hamiltonian for the SC lead $j$ and $H_{j,\rm{T}}$ defines the tunnel-coupling between the lead $j$ and the electronic level. The electrons residing in the QD level therefore obey
\begin{align}
	H_{\rm{d}}
	&=
	\varepsilon_{\rm{d}} \sum_{\sigma} 
	d^{\dagger}_{\sigma}
	d^{\vphantom{\dagger}}_{ \sigma}
	+
	U \,
	n_{\uparrow} n_{\downarrow},
	\label{eq:dotH}
\end{align}
with $d^{\dagger}_{\sigma}$ ($d^{\vphantom{\dagger}}_{\sigma}$) creating (annihilating) a dot electron of spin $\sigma \in \{\uparrow, \downarrow\}$ and $n_\sigma \equiv d^{\dagger}_{\sigma} d^{\vphantom{\dagger}}_{\sigma}$ representing the occupation number operator. Each SC follows a BCS Hamiltonian
\begin{align}
	H_{j,\rm{SC}}
	&=
	\sum_{\mathbf{k}\sigma} 
	\varepsilon_{\mathbf{k}j}
	c^{\dagger}_{\mathbf{k} j\sigma}
	c^{\vphantom{\dagger}}_{\mathbf{k} j\sigma}
	-
	\sum_{\mathbf{k}}
	\left(
	\Delta_{j}
	c^{\dagger}_{\mathbf{k}j \uparrow} 
	c^{\dagger}_{-\mathbf{k}j \downarrow}
	+
	\textit{H.c.}
	\right),
	\label{eq:bcsH}
\end{align}
while the coupling of the level to the SC lead $j$ obeys
\begin{align}
	H_{j,\rm{T}} 
	&=
	\sum_{\mathbf{k} \sigma}
	\left(
	V^*_{\mathbf{k}j}  
	c^{\dagger}_{\mathbf{k}j\sigma}
	d^{\vphantom{\dagger}}_{\sigma} 
	+ 
	V_{\mathbf{k}j}  
	d^{\dagger}_{\sigma}
	c^{\vphantom{\dagger}}_{\mathbf{k}j\sigma}
	\right),
	\label{eq:tunnelH}
\end{align}
where $c^{\dagger}_{\mathbf{k} j\sigma}$ ($c^{\vphantom{\dagger}}_{\mathbf{k}j\sigma}$) creates (annihilates) an electron of spin $\sigma$, quasi-momentum $\mathbf{k}$ and energy $\varepsilon_{\mathbf{k}j}$ in lead $j$. The leads couple to the QD only and not directly to each other.

All terminals are assumed to be made of the same SC material, reflecting the typical experimental scenario. The SC leads thus share the gap magnitude $\Delta$ and the bandwidth $2B$. Their BCS phases $\varphi_{j}$ on the other hand are individually controlled by flux or current bias. The coupling of the level to lead $j$ is characterized by the tunneling strength $\Gamma_j\equiv\pi\sum_{\mathbf{k}} |V_{\mathbf{k}j}|^2\delta(\omega-\varepsilon_{\mathbf{k}j})$, presumed energy independent for simplicity, with $\Gamma \equiv \sum_j \Gamma_j$ denoting the total hybridization of the level with the three leads. In what follows, the device geometry is conveniently expressed through the relative hybridization strengths $\gamma_j \equiv \Gamma_j/\Gamma$, which fulfill $\gamma_1+\gamma_2+\gamma_3=1$. Their configurations are thus fully defined by specifying only two of them, for which we choose $\gamma_2$ and $\gamma_3$ throughout the paper.

Following Refs.~\cite{Zalom-2024,Zalom-2026,Zalom-2026-arx}, we resolve the problem by solving for its Green's function. This is facilitated by the Hamiltonians~\eqref{eq:bcsH} and~\eqref{eq:tunnelH} being quadratic. Consequently, the SC environment can be integrated out exactly, and its entire effect on the QD level reduces to a corresponding self-energy contribution to the retarded QD Green's function. The latter is then used to resolve the full many-body problem, giving access to the spectrum, the ground state character, and the free energy. Applying the Nambu formalism with spinor $D^{\dagger} = \left(d^{\dagger}_{\uparrow}, d^{\vphantom{\dagger}}_{\downarrow}\right)$ allows to obtain the Green's function of the $d$-electrons as
\begin{align}
\mathbb{G}_{\rm{d}}(\omega)
=
\frac{1}{
		\omega \mathbb{1} 
		- \mathbb{H}_{\rm{d},0}
		- \mathbb{\Sigma}_{0}(\omega)
		- \mathbb{\Sigma}_U (\omega)
		+i \eta
		},
\label{eq:dotG}
\end{align} 
where boldface denotes $2 \times 2$ Nambu matrices, $\eta$ is infinitesimally small, and
\begin{align}
\mathbb{H}_{\rm{d},0}
=
\begin{pmatrix}
	\varepsilon_{\rm{d}} & 0
	\\
	0 & -\varepsilon_{\rm{d}}
\end{pmatrix}
\end{align}
is the Nambu representation of the non-interacting part of \eqref{eq:dotH}. The Coulomb interaction $U$ enters the Green's function via $\mathbb{\Sigma}_U(\omega)$, which needs to be obtained numerically. On the other hand, as already explained, the SC environment is quadratic and contributes $\mathbb{\Sigma}_0(\omega)$ hybridization self-energy which can be directly obtained by setting $U=0$ in~\eqref{eq:Hsingle}. A standard equation-of-motion technique then yields 
\begin{equation}
\mathbb{\Sigma}_0(\omega^+)
=	
\Gamma 
\begin{pmatrix} 	
	\omega & \Delta \bm{\chi}
	\\ 
	\Delta \bm{\chi}^*	& \omega 
\end{pmatrix}
F(\omega^+),
\label{eq:sigma}
\end{equation}
where $F(\omega^+)$ has a parametric dependence only on $\Delta$ and $B$. In the wide-band limit $B \to \infty$, it reads
\begin{equation}
	F(\omega^+)
	=
	\begin{cases}
		-\frac{1}{\sqrt{\Delta^2-\omega^2}},
		&
		\text{for }|\omega|<\Delta
		\\
		-\frac{i\, \mathrm{sgn}(\omega)}{\sqrt{\omega^2-\Delta^2}},
		&
		\text{for }|\omega|>\Delta.
	\end{cases}
\end{equation}
The off-diagonal elements of \eqref{eq:sigma} clearly depend on a single complex-valued and manifestly gauge-dependent quantity
\begin{equation}
	\bm{\chi}   
	\equiv     
	\sum_{j=1}^{3} 
	\gamma_j e^{i\varphi_j},
	\label{eq:chi}
\end{equation}
which is a sum of three phasors of lengths $\gamma_j$ and phases $\varphi_j$. The underlying gauge freedom allows rotating $\bm{\chi}$ onto the real axis via $\varphi_j \rightarrow \varphi_j -\text{Arg}(\bm{\chi})$. Such a global shift of all SC phases leaves the physics unchanged, as only phase differences are physically observable. This results in a fully gauge-invariant formulation of~\eqref{eq:sigma} which allows to replace in Eq.~\eqref{eq:sigma} the gauge dependent $\bm{\chi}$ with its gauge-independent amplitude
\begin{equation}
	\chi
	\equiv
	|\bm{\chi}|	
	=
	\sqrt{1-4\sum_{j>l=1}^3	\gamma_j\gamma_l \sin^2 \frac{\varphi_j-\varphi_l}{2}}.
	\label{eq:chi_sqrt}
\end{equation}
It only involves phase differences and hybridization strengths and is bounded by $0 \leq \chi \leq 1$.

Most importantly, the description via Eq.~\eqref{eq:chi_sqrt} extends beyond the non-interacting case. To this end, we notice that the SC environment enters the QD problem~\eqref{eq:dotG} exclusively through $\mathbb{\Sigma}_0(\omega)$, so the interaction self-energy $\mathbb{\Sigma}_U(\omega)$ knows nothing about the individual $\gamma_j$ and $\varphi_j$. Consequently, it is only $\chi$ that determines the full $\mathbb{G}_{\rm{d}}(\omega)$, the free energy $F(T)$ and all QD observables expressible via $\mathbb{G}_{\rm{d}}(\omega)$ once $U$, $\varepsilon_{\rm{d}}$, $\Gamma$ and $T$ are fixed. Consequently, two different configurations $\{\gamma_j, \varphi_j\}$ with equal $\chi$ possess identical many-body spectra, which allows to project any multi-terminal configuration onto a two-terminal one with equal couplings, provided the leads are not directly coupled, as first derived in Ref.~\cite{Zalom-2024}.

The geometric factor of such a two-terminal reference problem is $\chi=|\cos(\varphi/2)|$ with $\varphi$ its phase difference, so it covers the full range $0 \leq \chi \leq 1$ and thus all configurations of the present three-terminal device. It is additionally a numerically well tractable model that can be solved by standard Numerical Renormalization Group (NRG)~\cite{Satori-1992,Yoshioka-2000,Zalom-2023} or Quantum Monte Carlo (QMC)~\cite{Siano-2004,Luitz-2010,Luitz-2012} techniques in an unbiased way. For instance, for given $U$ and $\Gamma$, the two-terminal phase difference $\varphi^*$ at which a singlet-doublet transition occurs gives the corresponding critical $\chi^* = \cos(\varphi^*/2)$. Plotting isolines of $\chi=\chi^*$ from the analytic expression \eqref{eq:chi_sqrt} then yields the phase diagram of our device.

\section{Multi-terminal SC transistor effect \label{sec:sc_trans}}

\begin{figure*}[t]
	\includegraphics[width=2.0\columnwidth]{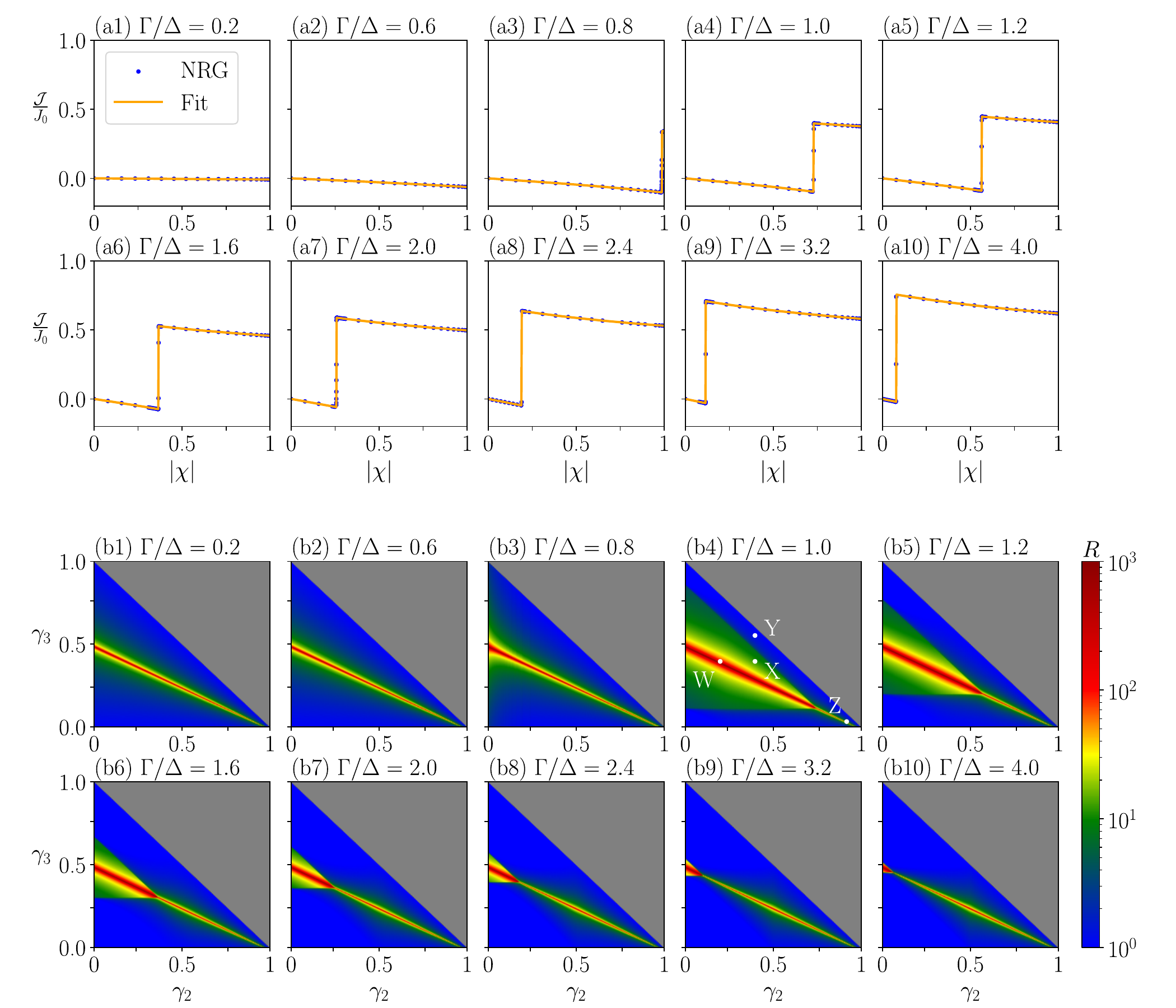}
	\caption{
		(a1)-(a10)
		Universal functions $\mathcal{J}_{T=0}(\chi)$ calculated from the coupling-symmetric two-terminal counterpart by employing NRG Ljubljana.
		(b1)-(b10)
		The switching ratio $R=J_{\rm{on}}/J_{\rm{off}}$ calculated from the corresponding values of critical currents $|J_2^c|$ in the corresponding state with fixed $\varphi_3$. Current calculations employ the mapping technique of~\cite{Zalom-2024}. Along the analytically obtained sweet line $\gamma_3=(1-\gamma_2)/2$, a region of significantly increased transistor yield is apparent. All maps show a triangular region of increased ratio $R$, which is, however, suppressed as $\Gamma$ increases.
		\label{fig:maps}}
\end{figure*}

To completely resolve the supercurrents in the three-terminal device, the formalism of the geometric factor $\chi$ represents the crucial tool. Following Ref.~\cite{Zalom-2024}, we calculate the supercurrent flowing from terminal $j$ to the QD by differentiating the total free energy $\mathcal{F}(T)$ with respect to $\varphi_j$ and using $\chi$ in the chain rule: 
\begin{align}
	J_j(T)
	=
	\frac{2e}{\hbar} 
	\frac{\partial \mathcal{F}(T)}{\partial\varphi_j}
	= 
	\frac{2e}{\hbar} 
	\frac{ \partial \mathcal{F}(T)}{\partial\chi}
	\frac{\partial \chi}{\partial \varphi_j} 
	= 
	\mathcal{J}_T(\chi)
	\chi'_j.
	\label{eq:current}
\end{align}
This defines two functions
\begin{align}
\mathcal{J}_T(\chi) 
&\equiv 
\frac{J_0}{\Delta}
\frac{\partial \mathcal{F}(T)}{\partial\chi},
\\
\chi'_j
&\equiv
\frac{\partial \chi}{\partial \varphi_j},
\end{align}
where $J_0 \equiv 2e\Delta/\hbar$ serves as the unit of supercurrent throughout the paper. Crucially, the prefactor $\mathcal{J}_T(\chi)$ depends only on $\chi$ while $\chi'_j$ is fully analytic and encodes the details about any $N$-terminal SC environment defined by its relative hybridizations $\gamma_j$ and phases $\varphi_j$. Consequently, at fixed $U$, $\varepsilon_{\rm{d}}$, $\Gamma$ and $T$ we can calculate $\mathcal{J}_T(\chi)$ from the simplest model that encompasses all values of $\chi$ in its allowed range of $[0,1]$. To this end the two-terminal coupling symmetric model is best suited. Using NRG or QMC, one can then determine the corresponding Josephson current $J(T)$ at phase difference $\varphi$, which appears on the left-hand side of Eq.~\eqref{eq:current}. In this system the geometric factor becomes $\chi = \cos(\varphi/2)$ yielding $\chi'=-\sin (\varphi/2)/2$ where we suppressed the lead indices in both the Josephson current and the derivative function. The universal prefactor is then extracted as
\begin{align}
	\mathcal{J}_T(\chi)
	=
	\frac{J(T)}{\chi'}.
	\label{eq:extraction}
\end{align}
The resulting functions $\mathcal{J}_T(\chi)$ are shown in Figs.~\ref{fig:maps}(a1)-(a10). They have been obtained by employing NRG Ljubljana~\cite{NRGzenodo} to the coupling-symmetric two-terminal model at zero temperature with $U/\Delta=3$ and $\Gamma$ varying accordingly. For such a parameter choice we can observe characteristic discontinuities at critical values $\chi^*$ [Figs.~\ref{fig:maps}(a4)-(a10)]. When $0 \leq \chi \leq \chi^*$ the universal function is negative yielding a characteristic $\pi$ supercurrent which signals an underlying doublet ground state. For $\chi^* \leq \chi \leq 1$ the universal function yields a characteristic $0$ supercurrent indicating a singlet ground state. Notably, there is a characteristic disparity between the considerably smaller magnitude of $\mathcal{J}_{T=0}(\chi)$ in the doublet regions and its maximum in the singlet ground state. This effect was exploited in Ref.~\cite{Zalom-2024} to design a supercurrent transistor effect that allowed for suitably
chosen $\gamma_j$ to drive a three-terminal device into a region of exclusively doublet ground state, where the output supercurrent was strongly suppressed.

The SC transistor effect of Ref.~\cite{Zalom-2024} relied thus on delicate tuning of $U$, $\Gamma$ and the relative hybridizations $\gamma_j$. It was exclusively facilitated by the presence of QPTs and would thus rapidly diminish at finite temperature. To overcome such severe limitations, we instead turn to the analytic factor that describes geometric details of the three-terminal device. In the gauge $\varphi_1=0$ we can use $\varphi_2$ and $\varphi_3$ directly as the gauge-invariant phase differences. Together with $\gamma_1+\gamma_2+\gamma_3=1$, this yields
\begin{align}
	\chi'_j
	=
	-\frac{\gamma_j}{\chi}
	\sum_{l \neq j}
	\gamma_l \sin (\varphi_j-\varphi_l),
	\label{eq:geometry_general}
\end{align}
where the geometric factor $\chi$ itself appears in the denominator. Specifically for the output terminal we obtain
\begin{align}
	\chi'_2
	=
	-\frac{\gamma_2}{\chi}
	\left[
	\gamma_1 \sin (\varphi_2)
	+
	\gamma_3 \sin (\varphi_2-\varphi_3)
	\right],
	\label{eq:geometry_J2}
\end{align}
which exhibits fundamentally different behavior as ${\varphi_3 = 0}$ and $\varphi_3 = \pm \pi$. For $\varphi_3=0$ we obtain
\begin{align}
	\left. \chi'_2 \right|_{\varphi_3=0 }
	= 
	-\frac{
		\gamma_2 (1- \gamma_2) \sin (\varphi_2)
	}{
		\chi
	}, 
	\label{eq:on}
\end{align}
whereas for $\varphi_3 = \pm \pi$
\begin{align}
	\left. \chi'_2 \right|_{\varphi_3=\pm \pi}
	=
	-\frac{
		\gamma_2 (\gamma_1 - \gamma_3) \sin (\varphi_2)
	}{
		\chi
	}.
	\label{eq:off}
\end{align}

Along the trajectory with $\varphi_3=0$, the geometric current factor vanishes identically only in the degenerate case $\gamma_2=1$, which corresponds to terminals $1$ and $3$ being completely disconnected from the system and is thus a trivial regime. In contrast, along the trajectory $\varphi_3=\pm \pi$ the numerator of \eqref{eq:off} vanishes identically when
\begin{align}
	\gamma_3
	=
	\frac{1-\gamma_2}{2}
	=
	\gamma_1,
	\label{eq:resonance}
\end{align}
leading to a perfectly closed supercurrent switch ($J_2=0$) via a gate-control of the hybridization strengths.

The origin of the effect is transparent at the level of the geometric factor $\bm{\chi}$ itself. Setting $\varphi_3=\pm\pi$ in \eqref{eq:chi} gives $\bm{\chi}=(\gamma_1-\gamma_3)+\gamma_2 e^{i\varphi_2}$. At $\gamma_1=\gamma_3$ the two phasors coming from the flux-loop cancel exactly leaving $\chi=\gamma_2$, which thus becomes independent of $\varphi_2$. The free energy of the QD in Eq.~\ref{eq:current}, being a function of $\chi$ alone, is then insensitive to the output phase. The output current $J_2$ must therefore vanish identically. At the same time, \eqref{eq:geometry_general} returns $\chi'_1=-\chi'_3=\gamma_1\gamma_2\sin(\varphi_2)/\chi \neq 0$ meaning that the corresponding supercurrents $J_1$ and $J_3$ are non-zero but fulfill $J_1=-J_3$. Consequently, the flux-loop sustains a purely circulating supercurrent while no charge is transmitted into the output lead. Such a scenario can be interpreted on a more general footing. By understanding the tunneling barriers between the electronic level residing in the QD region and leads $1$ and $3$ as individual Josephson junctions, we notice that the condition $\gamma_1=\gamma_3$ essentially requires them to be identical. In the closed state, the flux loop thus operates in a regime analogous to a dc SQUID with identical junctions at half a flux quantum. For negligible loop inductance, the two arm supercurrents of such a SQUID are exactly opposite at half flux, so that interference suppresses the transport critical current completely while a circulating current persists [Jaklevic-1964, Zimmerman-1966, Fulton-1972]. Junction asymmetry or finite loop inductance leaves a residual critical current [Fulton-1972, Tesche-1977, Koch-2007], in close analogy to the case $\gamma_1 \neq \gamma_3$ case of the present three-terminal device.

\section{Resonance width of the switching ratio \label{sec:resonance_widths}}

\begin{figure*}[!htbp]
	\includegraphics[width=2.1\columnwidth]{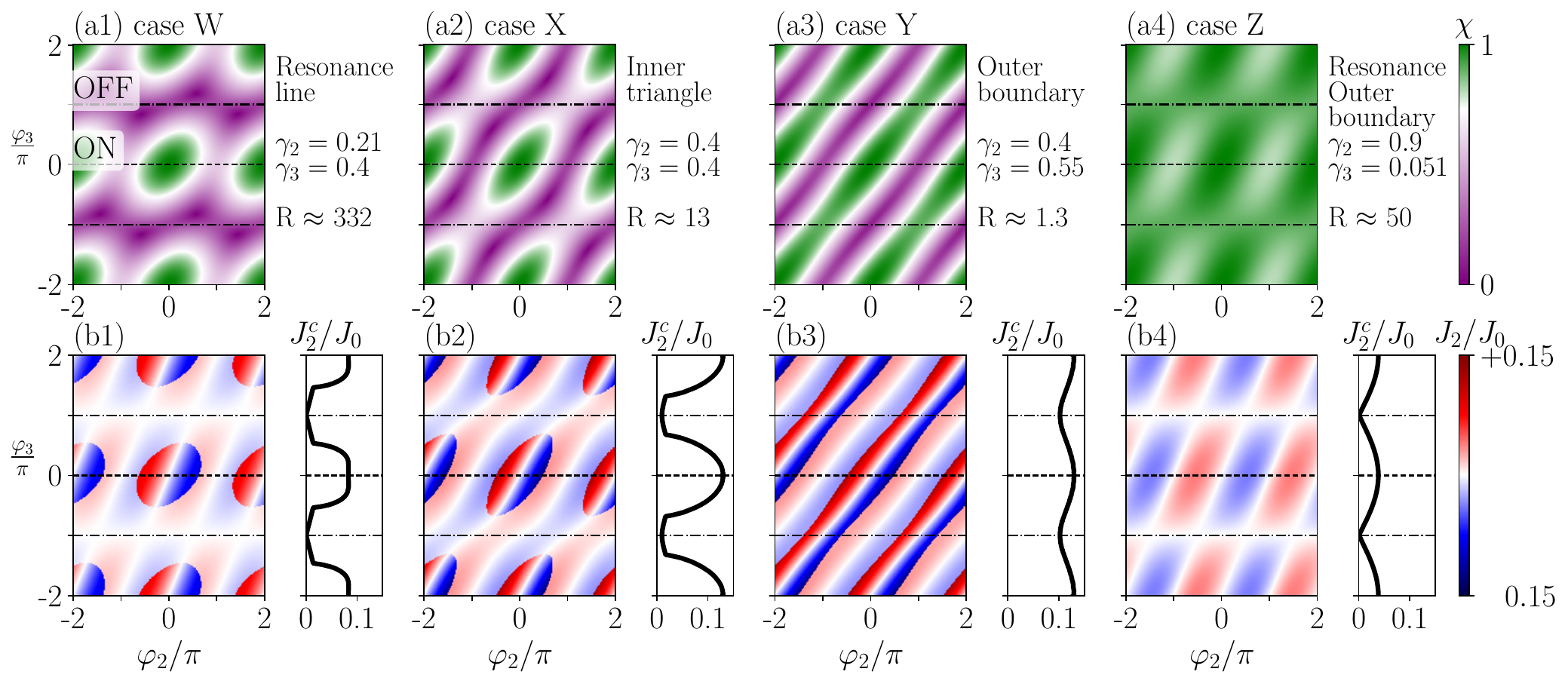}
	\caption{
		(a1)-(a4)
		Phase diagrams in the $\varphi_2$-$\varphi_3$ plane for cases $\rm{W}$, $\rm{X}$, $\rm{Y}$ and $\rm{Z}$. Only singlet (green color) and doublet (violet) ground states occur. Stripes and islands occur in the phase diagrams in accordance with~\cite{Zalom-2024}.
		(b1)-(b4)
		The corresponding $\varphi_2$-$\varphi_3$ maps of the $J_2$ currents corresponding to panels (a1)-(a4). The insets to the right of each panel show the magnitude of critical currents $|J_2^c|$ for a fixed value of $\varphi_2$.
	   \label{fig:examples}
       }
\end{figure*}

The multi-terminal SC transistor effect can thus be understood in purely geometric terms with phases being operated in-situ by flux- or current drive and tunnel barriers by gate-voltages. While the former switch the device between open and closed states, the latter is required to tune to the ideal closed regime given by $\gamma_1=\gamma_3$ condition of Eq.~\eqref{eq:resonance}. Since $\gamma_2=1-\gamma_1-\gamma_3$, it is natural to select relative hybridizations $\gamma_2$ and $\gamma_3$ for characterization of the device. The parameter $\gamma_2$ is expected to directly influence the output supercurrent $J_2$ while $\gamma_3$ governs directly supercurrents in the loop. In the $\gamma_2$-$\gamma_3$ parametric plane, condition \eqref{eq:resonance} defines then a line $\gamma_3=(1-\gamma_2)/2$ along which the three-terminal device switches between perfectly closed ($\varphi_3=\pm \pi$) and open ($\varphi_3=0$) states. 

Since the output supercurrent in the open state (denoted as $J_{\mathrm{on}}$) is finite but its counterpart in the closed state ($J_{\mathrm{off}}$) vanishes, their ratio defining the switching ratio diverges, i. e. $R=J_{\mathrm{on}}/J_{\mathrm{off}} \rightarrow \infty$. Eq.~\eqref{eq:resonance} acts thus as a resonance condition for tuning the couplings to obtain the ideal SC transistor characterized by $R \rightarrow \infty$. However, exact control over tunneling barriers in experiments is challenging to achieve. The central question is thus how broad is the resonance $R \rightarrow \infty$ with regard to tuning of the tunneling barriers. 

To this end, we turn our attention to the detailed interaction effects encoded in $\mathcal{J}_T(\chi)$, which was obtained numerically as explained in Sec.~\ref{sec:sc_trans} with results plotted in Figs.~\ref{fig:maps}(a1)-(a10). The model parameters are selected as $U/\Delta=3$, while the central QD is set to half-filling, i.~e. $\varepsilon_{\rm{d}}=-U/2$. Then we vary the ratio $U/\Gamma$. The output supercurrent in the three-terminal scenario is then obtained by multiplying the numerically obtained $\mathcal{J}_{T=0}(\chi)$ function with the proper geometric current factor $\chi'_2$. This way, we can generate $J_2$ currents at any $\gamma_2$-$\gamma_3$ configuration of the three-terminal device at given $U$, $\varepsilon_{\rm{d}}$ and $\Gamma$. From the $\varphi_2$ dependence of $J_2$ at $\varphi_3=0$ and $\varphi_3=\pm \pi$ one then reads off the critical values of supercurrent corresponding to the maximum allowed supercurrent $J_{\mathrm{on}}$ in the open state and the corresponding maximum allowed supercurrent $J_{\mathrm{off}}$ in the closed state. The resulting switching ratio $R$ is then shown in Figs.~\ref{fig:maps}(b1)-(b10) for the triangular regions of allowed $\gamma_2$ and $\gamma_3$ strengths.

In all selected cases of $\Gamma/\Delta$, dark-red lines emerge at the position predicted by \eqref{eq:resonance}. Although sweet lines are present regardless of exact model parameters, the behavior of $R$ around their position shows complex behavior due to $\Gamma$-$U$ interplay. For $\Gamma/\Delta < 1$ [Figs.~\ref{fig:maps}(b1)-(b3)], the resonances are sharp and their width decreases monotonically as $\gamma_2 \rightarrow 1$. At $\Gamma=\Delta$ [Fig.~\ref{fig:maps}(b4)], bimodal behavior emerges with a broad resonance for $\gamma_2 < 0.65$ that abruptly switches to a significantly narrower width at $\gamma_2 \sim 0.65$ until $\gamma_2 \rightarrow 1$. The broad nature is associated with a green-shaded inner triangle region where the switching ratio is significant. Sharp resonance occurs in the outer dark blue-shaded region with $R \approx 1$ that shows only irrelevant transistor effect outside of the resonance condition. Increasing $\Gamma$ expands this region irrelevant for supercurrent switching at the expense of the inner triangle. At $\Gamma=4\Delta$ [Fig.~\ref{fig:maps}(b10)] the remaining inner triangle region has almost entirely vanished. 

Consequently, Figs.~\ref{fig:maps}(b1)-(b10) clearly define the operating regime of the device. The sweet line itself is present throughout, but only at intermediate hybridization, $\Gamma \approx \Delta$, is it surrounded by a broad region of elevated $R$. Both weaker and stronger coupling shrink this region: at small $\Gamma$ [Figs.~\ref{fig:maps}(b1)-(b3)] the resonance appears as a thin line, while at large $\Gamma$ the broad resonant feature in the green-shaded triangle is progressively diminished in favor of the outer region where $R \approx 1$. Operating the device at intermediate $\Gamma$ therefore allows one to exploit the broad resonances for an effective tuning of the switching ratio. Tuning of the $\gamma_2$ values, on the other hand, matters only with respect to the bimodal behavior, and a proper value shall only ensure that the device is operated within the green-shaded area.

\subsection{Origin of the resonance-width structures \label{subsec:origin}}

We now analyze the origin of the structures observed in the switching-ratio maps. The geometric current factor $\chi'_2$ clearly defines the resonance line throughout the entire $\gamma_2$-$\gamma_3$ plane while the universal function $\mathcal{J}_T(\chi)$ serves as a modulating prefactor. Its role will thus be crucial especially away from the resonance line given by~Eq.~\ref{eq:resonance}. Consequently, both the inner triangle of elevated $R$ and the outer region with $R \approx 1$, as observed for example in~Fig.~\ref{fig:maps}(b4), must stem from the singlet-doublet QPTs encoded in $\mathcal{J}_T(\chi)$. 

We therefore select the $\Gamma/\Delta=1$ case where these distinct regions of switching ratio are all co-existing and select four representative points $\rm{W}$-$\rm{Z}$ according to Fig.~\ref{fig:maps}(b4). $\rm{W}$ lies in the green-shaded inner triangle and close to the sweet line, $\rm{X}$ lies within the green-shaded inner triangle but does not fulfill~\eqref{eq:resonance}, $\rm{Y}$ lies in the outer dark-blue boundary away from the resonance~\eqref{eq:resonance} while $\rm{Z}$ lies close to the resonance condition but still in the same region. For this discussion, we again choose the gauge $\varphi_1=0$, so values of $\varphi_2$ and $\varphi_3$ directly represent the gauge-independent phase differences of our device. Their variation shows singlet (green) and doublet (violet) ground states emerging in phase diagrams~\ref{fig:examples}(a1)-(a4). These are obtained from the geometric factor $\chi$ with numeric input of $\chi^*=0.721$ from the corresponding NRG calculation. Using additionally $\mathcal{J}_{T=0}$ obtained from the same NRG calculation, the corresponding $J_2$ supercurrents are derived in Figs.~\ref{fig:examples}(b1)-(b4). The side panels display $J_2^c$ as a function of the $\varphi_3$-configuration of the device.

\begin{figure*}[!htbp]
	\includegraphics[width=2.1\columnwidth]{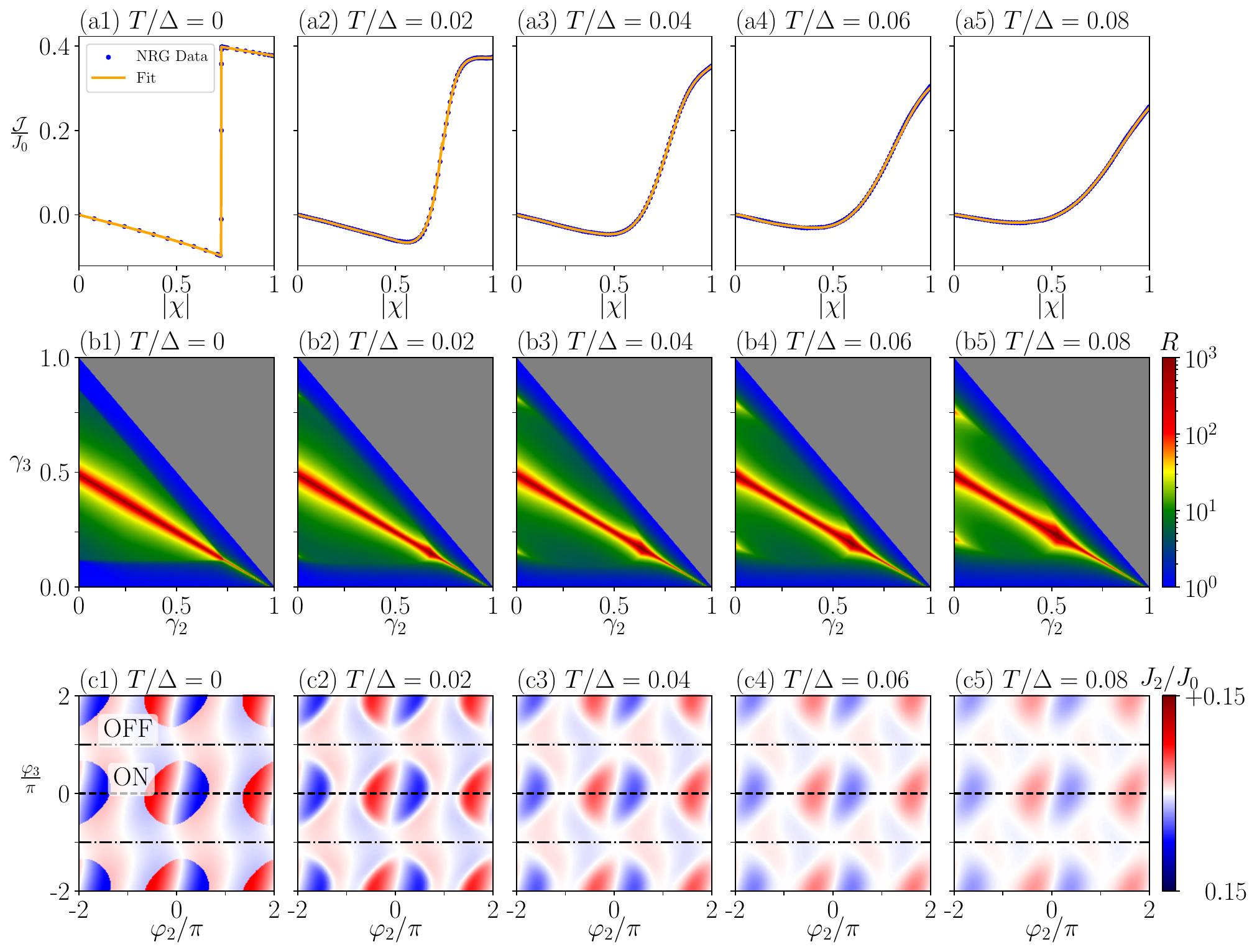}
	\caption{
		  (a1)-(a5) 
        Universal functions $\mathcal{J}_T(\chi)$ obtained from NRG Ljubljana at temperatures $T/\Delta \in \{0, 0.02, 0.04, 0.06, 0.08\}$ (panels left to right). Other model parameters are selected according to Fig.~\ref{fig:maps}(a4): $U/\Delta=3$, $\varepsilon_{\rm{d}}=-U/2$ and $\Gamma/\Delta=1.0$. The discrete jump present at $T=0$ progressively smoothens into a continuous
		crossover and the overall amplitude decreases with increasing $T/\Delta$.
		(b1)-(b5) 
        Switching ratio $R=J_\mathrm{on}/J_\mathrm{off}$ across the $\gamma_2$-$\gamma_3$ parameter space. The sweet line and the surrounding region of significant transistor yield remain visible up to the highest temperature shown. The resonance width remains stable with only a minor decrease in the width throughout the temperature range shown.
		(c1)-(c5) 
        Current-phase maps of $J_2$ at $\gamma_2=0.2$ and $\gamma_3=0.3$, showing the open (ON) and closed (OFF) regimes. With increasing $T/\Delta$, an overall reduction in the magnitude of supercurrents is observed. Since this affects all regimes, the switching ratio $R$ remains stable.
	   \label{fig:temperature}
       }
\end{figure*}

Comparing the first cases $\rm{W}$ and $\rm{X}$, we notice that the overall structure of singlet islands immersed in a continuum of doublet ground states is preserved. Nevertheless, the geometrically dependent switching mechanism crucially alters the $J_2$ profile in the doublet continuum. Being close to the resonance condition, case $\rm{W}$ develops a horizontally aligned area of strongly suppressed $J_2$ along $\varphi_3=\pm\pi$. These are absent in $\rm{X}$, as a geometrically-dependent suppression mechanism is unavailable because of the distance from the resonance condition \eqref{eq:resonance}. As the inset of Fig.~\ref{fig:examples}(b2) shows, $J_2^c$ still reaches its minima at $\varphi_3 = \pm\pi$, but unlike in $\rm{W}$, the output supercurrent remains small but still non-zero. As a consequence, the transistor switching ratio does not increase without bound and instead reaches $\approx 13$. 

While the involvement of the geometric mechanism is superior to the one relying solely on the QPT mechanism of Ref.~\cite{Zalom-2024}, for supercurrent transistor switching, the latter is sufficient to generate the distinct triangular-shaped regions of ratio $R>1$ (green-shaded inner triangles) as observed in Fig.~\ref{fig:maps}. This pattern is tightly connected to the underlying phase diagram in the $\varphi_2$-$\varphi_3$ with singlet islands (green) being well separated from each other by broad regions of doublet phase when $\varphi_3$ is tuned. 

However, once such a structure is broken, as in the case $\rm{Y}$, $\varphi_2$ tunes only between alternating singlet and doublet stripes regardless of $\varphi_3$. Consequently, only an insignificant variation in $J_2^c$ is observed as shown in the inset of Fig.~\ref{fig:examples}(b3). The resulting transistor switching ratio drops to $\approx 1.3$. In general, striped phase diagrams are associated with the outer regions of the suppressed transistor ratio as previously observed in Fig.~\ref{fig:maps}. Unlike in the inner triangular regions, strong interactions and the underlying quantum phases form no suitable structures to generate the transistor effect.
	
As the case $\rm{Z}$ demonstrates, the geometric mechanism generating supercurrent switching is still present and leads to the observed thin resonances around the condition \eqref{eq:resonance}. As shown in the inset of Fig.~\ref{fig:examples}(b4), the critical $J_2^c$ current is still relatively large in the open regime of the three-terminal device when compared to cases $\rm{W}$ and $\rm{X}$. However, Fig.~\ref{fig:examples}(b4) also clearly demonstrates that zero-current regions are extremely narrow when tuned via $\varphi_3$, which is detrimental for gate-tuning in real devices.

\subsection{Temperature dependence of the resonance-width structures \label{subsec:temperature}}

The structures identified in Sec.~\ref{subsec:origin} rest on the discontinuities in $\mathcal{J}_{T=0}(\chi)$ that stem from singlet-doublet transitions. These become smoothed out at any finite temperature, which inevitably affects the coupling dependencies of the switching ratio. Nevertheless, the geometric current factor $\chi'_2$ is temperature independent, so the resonance condition \eqref{eq:resonance} itself cannot be affected. What remains to be examined are thus the regions of elevated $R$ surrounding it.

Let us therefore demonstrate such effects, focusing on the case of a sizable resonance width provided by $\Gamma/\Delta=1.0$, as evaluated in Fig.~\ref{fig:maps}(b4) and select a range of temperatures up to $T/\Delta=0.08$. First the corresponding $\mathcal{J}_T$ functions are obtained as shown in Figs.~\ref{fig:temperature}(a1)--(a5). The calculations have been performed as previously by employing the spectrally equivalent coupling-symmetric SC-AIM using the full density-matrix numerical renormalization group approach implemented within NRG Ljubljana.

At $T=0$, the $\mathcal{J}_{T=0}$ function possesses a discrete jump coinciding with the boundaries of the $0$ and $\pi$ phases. As shown in Figs.~\ref{fig:temperature}(a2)--(a5), this discontinuity disappears once $T>0$.
The supercurrent jump is instead smoothed out while simultaneously the amplitude of $\mathcal{J}_T$ decreases. This significantly reduces the critical supercurrents both in the $0$ and $\pi$ phases.

Using \eqref{eq:current}, the resulting dependencies of the switching ratio on the relative couplings are then obtained as shown in Figs.~\ref{fig:temperature}(b1)-(b5). The overall reduction of critical supercurrents is seen to have only a relatively minor impact on the resonance width in the selected temperature range, which is due to simultaneous reduction of critical currents in both the open and closed regimes of the device, as can be inferred from Figs.~\ref{fig:temperature}(c1)--(c5). From the practical perspective, temperature is therefore not detrimental to the switching ratio. Nevertheless, the overall suppression of critical supercurrents diminishes the magnitude of supercurrents transmitted by the device, which sets a practical lower bound on the operating temperature for applications requiring a minimum current throughput.

\section{Beyond the single-level scenario \label{sec:zbw}}

\begin{figure*}[t]
\includegraphics[width=2.0\columnwidth]{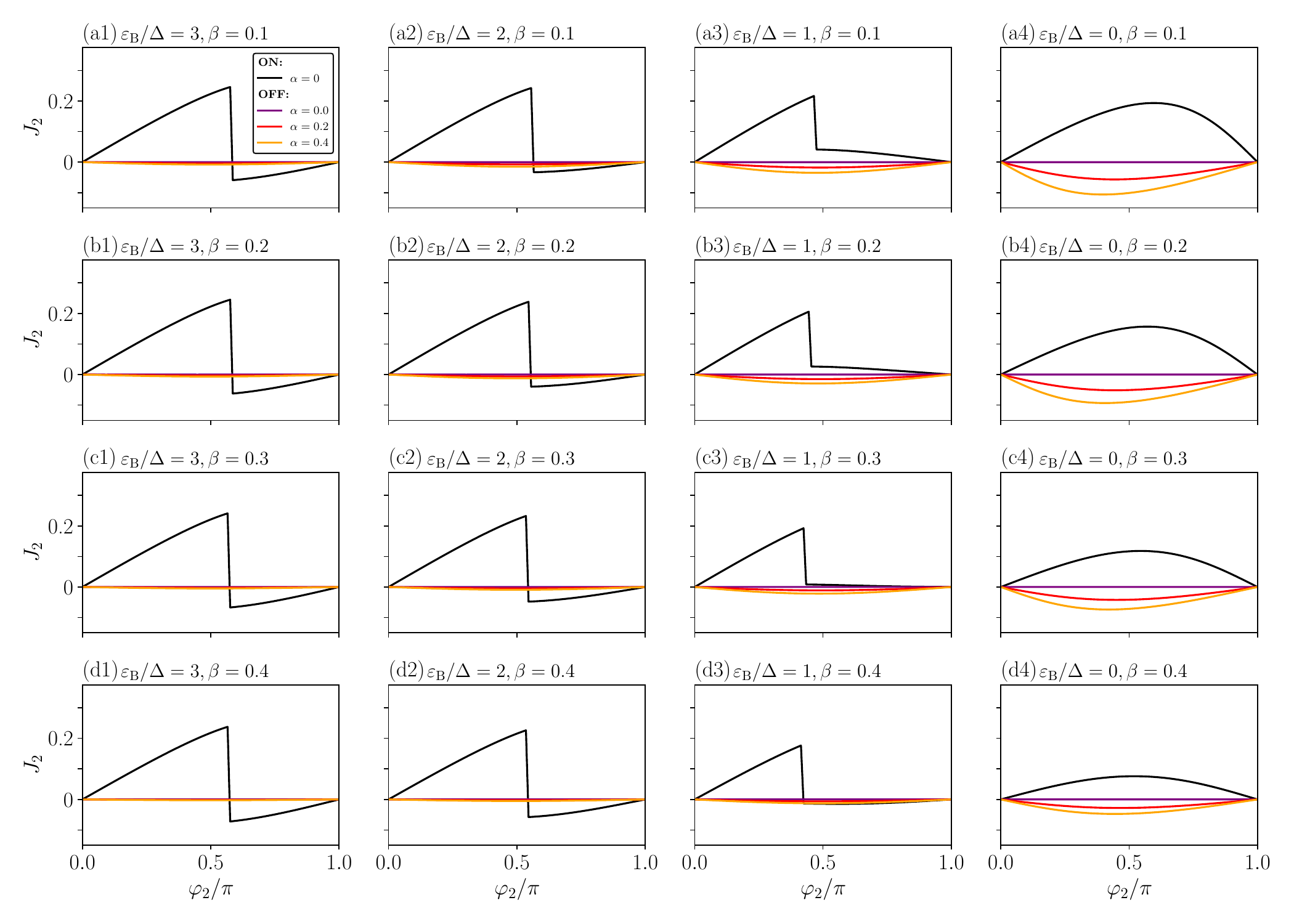}
	\caption{
	Current-phase relation of the output supercurrent $J_2$ in the three-terminal structure with two energy levels $\varepsilon_{\rm{A}}$ and $\varepsilon_{\rm{B}}$. The level A is kept at half-filling by setting $\varepsilon_{\rm{A}}/\Delta=-U_{\rm{A}}/(2\Delta)=-1.5$ ($U_{\rm{A}}/\Delta=3$), while $U_{\rm{B}}/\Delta=2$ and $\varepsilon_{\rm{B}}$ is decreased from $\varepsilon_{\rm{B}}/\Delta=3$ (leftmost column) towards $\varepsilon_{\rm{B}}/\Delta=0$ (rightmost column). Each panel shows $J_2(\varphi_2)$ in the open regime (black) as well as in the closed regime at $g_{\rm{1B}}$-$g_{\rm{3B}}$ asymmetry values $\alpha \in \{0, 0.2, 0.4\}$ (blue, orange, red)
    and $t/\Delta=0.6$. Closing of the gap between the two levels reshapes $J_2(\varphi_2)$ by removing the $0$-$\pi$ transition observed when the two levels are most separated, which keeps level A dominant. At $\varepsilon_{\rm{B}}=0$ both levels contribute comparably. This is then detrimental to the supercurrent transistor switching ratio.	
    \label{fig:zbw}}
\end{figure*} 

The switching mechanism established in Secs.~\ref{sec:sc_trans} and~\ref{sec:resonance_widths} rests on the presence of a single dominant level in the QD region. Realistic devices, however, host additional levels, and although these in principle can be pushed away in energy by electrostatic gating, they typically cannot be removed from the transport window entirely. To assess their influence on the SC transistor effect, we introduce a second level to the QD region as a minimal extension of~\eqref{eq:Hsingle}. 

We designate label $\rm{A}$ for the dominant level, which in 2DEG-based devices is typically the last one to hold electrons before the central region is fully depleted electrostatically. The nearest level in the hierarchy is then denoted by label $\rm{B}$. We assume that both levels couple to the three SC leads, which keep their labels $j \in \{ 1,2,3\}$ according to Fig.~\ref{fig:sc_trans}. Crucially, while barriers between the leads and the QD region are gate-tunable they cannot be individually adjusted with respect to the levels $\rm{A}$ and $\rm{B}$. Moreover, the orbital structure of level $\rm{B}$ typically differs from that of $\rm{A}$. Therefore, a barrier configuration that equalizes tunnelings between the leads forming the flux-loop and level $\rm{A}$, generically leaves the corresponding tunneling rates for level $\rm{B}$ asymmetric. We can thus realistically not expect that the geometric mechanism would suppress transport through both levels simultaneously in the closed regime, which results in inevitable leakage currents through the output terminal $2$.  

To qualitatively understand which effects contribute the most to the leakage, we employ the zero-bandwidth (ZBW) approximation for the SC leads. Technically, we thus replace all of the $\bf{k}$ modes appearing in Eq.~\eqref{eq:bcsH} by a single effective mode whose off-diagonal pairing is $\Delta e^{i\varphi}$ with $\varphi$ being the original BCS phase and $\Delta$ the bare gap parameter. The ZBW description involving two levels coupled to three SC leads then reads 
\begin{align}
H_{\mathrm{ZBW}}
\,=\,
H_{\rm{A}} + H_{\rm{B}} 
\,+\,
\sum_{j=1,2,3} 
\left(
H_{j,\rm{SC}}
\,+\,
H_{j,\rm{A}}^t
\,+\,
H_{j,\rm{B}}^t
\right),
\end{align}
where the levels A and B of energies $\varepsilon_{\rm{A}}$ and $\varepsilon_{\rm{B}}$ follow
\begin{align}
H_{\rm{A}} 
&=
\sum_{\sigma}   
\varepsilon_{\rm{A}}
d^\dagger_{\sigma,\rm{A}}
d^{\hphantom{\dagger}}_{\sigma,\rm{A}}
+
U_{\rm{A}} 
d^\dagger_{\uparrow,\rm{A}}
d^{\hphantom{\dagger}}_{\uparrow,\rm{A}}
d^\dagger_{\downarrow,\rm{A}}
d^{\hphantom{\dagger}}_{\downarrow,\rm{A}},
\\
H_{\rm{B}} 
&=
\sum_{\sigma}   
\varepsilon_{\rm{B}}
d^\dagger_{\sigma,\rm{B}}
d^{\hphantom{\dagger}}_{\sigma,\rm{B}}
+
U_{\rm{B}}
d^\dagger_{\uparrow,\rm{B}}
d^{\hphantom{\dagger}}_{\uparrow,\rm{B}}
d^\dagger_{\downarrow,\rm{B}}
d^{\hphantom{\dagger}}_{\downarrow,\rm{B}}
\label{eq:zbw}
\end{align}
with $d^\dagger_{\sigma,m}$ ($d^{\hphantom{\dagger}}_{\sigma,m}$) creating (annihilating) a dot electron of spin $\sigma$ on level $m \in \{ \rm{A},\rm{B} \}$. Coulomb repulsion terms $U_{\rm{A}}$ and $U_{\rm{B}}$ are generally not equal and represent material properties of a given device. The ZBW Hamiltonian for lead $j \in \{1,2,3\}$ contains the Hamiltonian for the three spinful modes created via $c^\dagger_{j\sigma}$, which obey
\begin{align}
H_{j,\rm{SC}}
&=
\Delta 
\left(
e^{i\varphi_j}
c^\dagger_{j\uparrow}
c^\dagger_{j\downarrow}
+
e^{-i\varphi_j}
c^{\hphantom{\dagger}}_{j\downarrow}
c^{\hphantom{\dagger}}_{j\uparrow}
\right)
\label{eq:ZBW_BCS}
\end{align}
with the gauge $\varphi_1=0$ selected throughout this section. Six couplings $g_{j\rm{A}}$ and $g_{j\rm{B}}$ with $j \in \{1,2,3\}$ describe how leads~\eqref{eq:ZBW_BCS} couple the two levels via
\begin{align}
H_{j,\rm{A}}^t
&=
g_{j\rm{A}}
\sum_{\sigma}
d^\dagger_{\sigma,\rm{A}}
c^{\hphantom{\dagger}}_{j\sigma}
+
h.c.,
\\
H_{j,\rm{B}}^t
&=
g_{j\rm{B}}
\sum_{\sigma}
d^\dagger_{\sigma,\rm{B}}
c^{\hphantom{\dagger}}_{j\sigma}
+
h.c.,
\end{align}
where the coupling constants $g_{j\rm{A}}$ and $g_{j\rm{B}}$ should be understood as phenomenological quantities which are only indirectly related to the microscopic tunneling rates. 

Since we aim to characterize leakage processes implicated by the coupling-asymmetries of the additional level B, we subject level A to the resonant condition~\eqref{eq:resonance} by demanding $g_{1\rm{A}} = g_{3\rm{A}}$. Moreover, we select the center of a triangle-shaped region of the $\gamma_2$-$\gamma_3$ plane by setting $g_{1\rm{A}} = g_{2\rm{A}} = g_{3\rm{A}} = g$. The total tunneling rate of level A in the ZBW approximation therefore reads
\begin{align}
\Gamma_{\rm{A},\mathrm{ZBW}}
\equiv
\sum_j
\pi \frac{(g_{j\rm{A}})^2}{2}
=
\frac{3\pi g^2}{2}.
\end{align}

Subsequently, we parametrize two types of coupling asymmetries of level $\rm{B}$ via the parameters $\alpha$ and $\beta$ via
\begin{align}
g_{1\rm{B}}
&=
\sqrt{1-\alpha}
\sqrt{1+\beta}
\, t,
\\
g_{2\rm{B}}
&=
\sqrt{1-2\beta}
\, t,
\\
g_{3\rm{B}}
&=
\sqrt{1+\alpha}
\sqrt{1+\beta}
\, t.
\end{align}
Clearly, $\alpha=\beta=0$ corresponds to the totally symmetric coupling of the level $\rm{B}$ to the SC modes of the leads with total hybridization fulfilling
\begin{align}
\Gamma_{\mathrm{B, ZBW}}
\equiv
\sum_j
\pi \frac{(g_{j\rm{B}})^2}{2}
=
\frac{3\pi t^2}{2}
\end{align}
at arbitrary $\alpha,\beta \leq 1$. The effect of $\alpha$ is to introduce coupling asymmetry with respect to terminals $1$ and $3$, which is thus primarily responsible for the leakage through level $\rm{B}$. On the other hand, increasing $\beta$ progressively diminishes the coupling of the output SC lead to level B. The states of the transistor are not further optimized and are selected with reference to the ideal single level scenario. In other words,  $\varphi_3=0$ ($\varphi_3=\pi$) are per definition considered as the open (closed) states of the two-level system. From the resulting ground state energy $E_{\mathrm{GS}}$ we then obtain $J_2(\varphi_2)$ via \eqref{eq:current} as $J_2(\varphi_2)= (2e/\hbar) \partial E_{\mathrm{GS}}/\partial \varphi_2$. The obtained current-phase relations $J_2(\varphi_2)$ are then evaluated at different asymmetry parameters $\alpha$ and $\beta$ with respect to relative energies of two levels and their possible inter-level Coulomb interactions.

\subsection{Leakage current due to the proximity of additional level \label{subsec:proximity} }

We first investigate the leakage with respect to the mutual position of levels $\rm{A}$ and $\rm{B}$. The results for $J_2$ are shown in Fig.~\ref{fig:zbw}. The first column [Figs.~\ref{fig:zbw}(a1)-(d1)] is obtained for a sizable separation of the two levels by selecting $\varepsilon_{\rm{A}}/\Delta=-U_{\rm{A}}/2=-1.5$ and $\varepsilon_{\rm{B}}/\Delta=3$. Level $\rm{A}$ resides thus at the half-filled point which is a condition we keep throughout this section. Due to the sizable distance, level $\rm{B}$ interferes with the transport via $\rm{A}$ only in a marginal way. The supercurrent $J_{2,\mathrm{on}}(\varphi_2)$ in the open regime shows a clear $0$-$\pi$ transition point at essentially the same $\varphi_2/\pi=0.6$ for all selected values of $\alpha \in \{ 0.0, 0.2, 0.4 \}$ and $\beta \in \{ 0.1, 0.2, 0.3, 0.4 \}$. Since the current-phase relations $J_{2,\mathrm{off}}(\varphi_2)$ in the closed regime is almost constantly zero, the two-level system can be considered practically closed. 

The energy of level $\varepsilon_{\rm{B}}$ approaches then progressively that of level $\varepsilon_{\rm{A}}$ in the subsequent columns of Fig.~\ref{fig:zbw}. As $\varepsilon_{\rm{B}} \rightarrow 0$, the detrimental effects on the ideal switching mechanism become more pronounced, for two readily identifiable reasons. First, the output current $J_{2,\mathrm{on}}(\varphi_2)$ in the open regime becomes increasingly reshaped resulting in a diminishing of the critical current $J_{2,\mathrm{on}}$ as $\varepsilon_{\rm{B}} \rightarrow 0$. Second, level $\rm{B}$ leaks increasingly larger supercurrents in the closed regime with Figs.~\ref{fig:zbw}(a4)-(d4) showing scenarios with comparable critical values of supercurrents in open and closed regimes. The critical current $J_{2,\mathrm{off}}$ then becomes comparable to its open-state counterpart $J_{2,\mathrm{on}}$, reducing thus the supercurrent switching ratio towards impractical values.

Crucially, we observe two main trends when changing $\alpha$ and $\beta$. Increasing the loop coupling-asymmetry $\alpha$, which breaks the geometric resonance condition $g_{\rm{1B}} = g_{\rm{3B}}$ for level $\rm{B}$, raises the leakage current $J_{2,\mathrm{off}}$ in the closed state. On the other hand, increasing $\beta$, which reduces the coupling $g_{\rm{2B}}$ of level $\rm{B}$ to the output terminal $2$, suppresses the leakage current. While the former is detrimental for supercurrent switching, the latter has benign consequences for the design and operation of the three-terminal device. Consequently, by raising the barrier between the QD region and terminal $2$, both $g_{\rm{2A}}$ and $g_{\rm{2B}}$ are expected to decrease in comparison to the loop coupling strengths, which then drives $\beta$ effectively towards its optimal operation value. Such an approach is also consistent with the results on a single dominant level $\rm{A}$ as obtained in Fig.~\ref{fig:maps}, where relative tunneling rates of $\gamma_2$ approaching $0$ generally yield a wide resonance in $R$, making the SC transistor effect more accessible in practical terms.

\begin{figure*}[t]
\includegraphics[width=2.0\columnwidth]{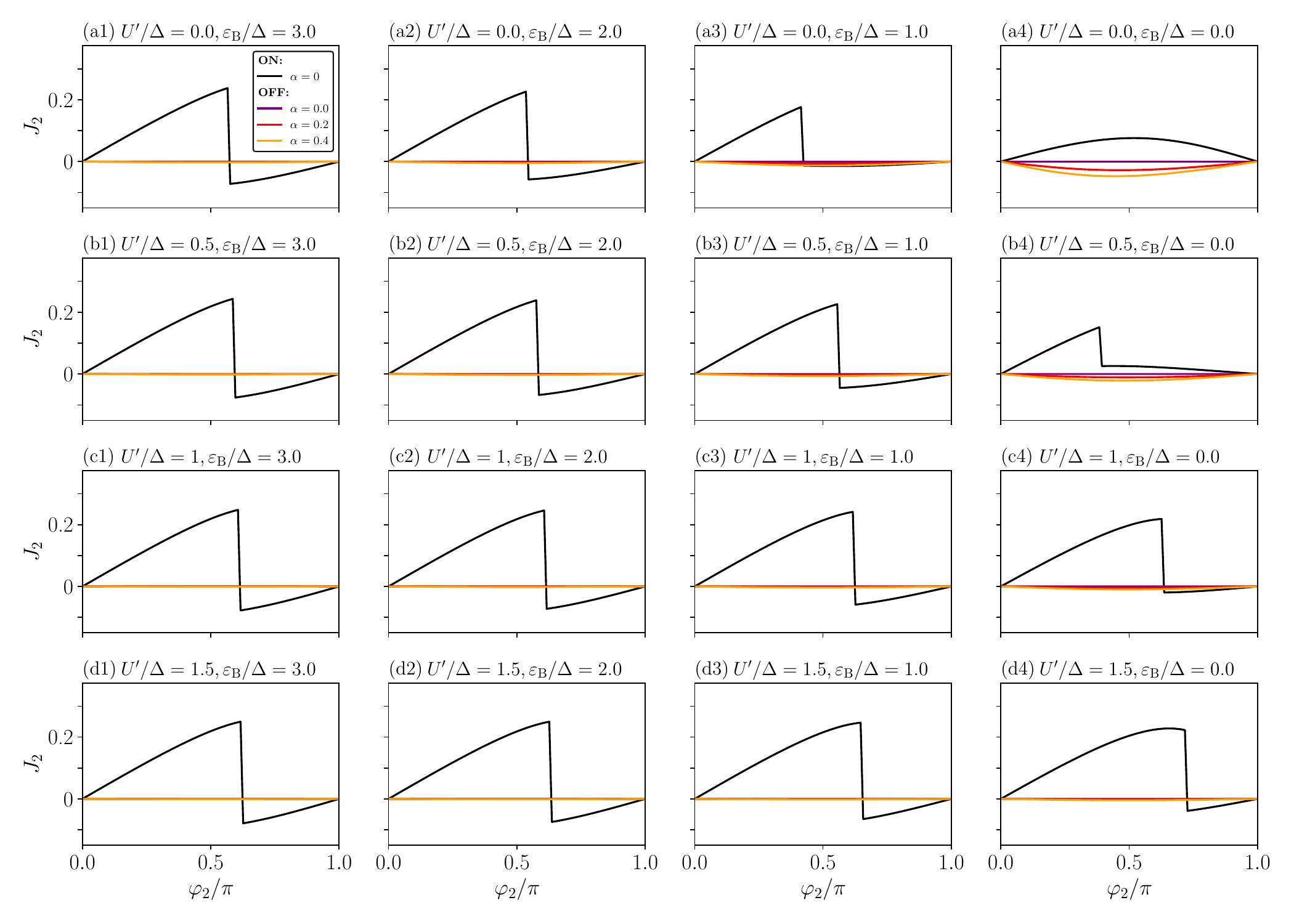}
	\caption{
    Current-phase relation of the output supercurrent $J_2$ in the three-terminal structure with two energy levels $\varepsilon_{\rm{A}}$ and $\varepsilon_{\rm{B}}$ in the presence of the inter-level Coulomb repulsion $U'$. The level $\rm{A}$ is kept at half-filling by setting $\varepsilon_{\rm{A}}/\Delta=-U_{\rm{A}}/2=-1.5$ ($U_{\rm{A}}/\Delta=3$), while $U_{\rm{B}}/\Delta=2$, $t/\Delta=0.6$ and $\beta=0.4$. The level separation decreases from left to right as $\varepsilon_{\rm{B}}/\Delta=3$ (leftmost column) towards $\varepsilon_{\rm{B}}/\Delta=0$ (rightmost column), whereas $U'$ is fixed within each row and increases downwards from $U'/\Delta=0$ [(a1)-(a4)] to $U'/\Delta=1.5$ [(d1)-(d4)]. Each panel shows $J_2(\varphi_2)$ in the open regime (black) as well as in the closed regime at $g_{\rm{1B}}$-$g_{\rm{3B}}$ asymmetry values $\alpha \in \{0, 0.2, 0.4\}$ (blue, orange, red). The uppermost row reproduces the $\beta=0.4$ results of Fig.~\ref{fig:zbw}. Increasing $U'$ shifts level $\rm{B}$ upwards and thereby restores the current-phase relation characteristic of well-separated levels, recovering the $0$-$\pi$ transition and suppressing the leakage in the closed regime.
    \label{fig:Uinter}
    }
\end{figure*} 

\subsection{Leakage current due to the inter-level Coulomb repulsion \label{subsec:inter} }

Inter-level Coulomb repulsion $U'$ represents a ubiquitous phenomenon in systems where multiple levels are embedded in Josephson junctions. It can severely alter both the ABS spectrum and the resulting transport properties~\cite{Logan-2014,Wrzesniewski2017-kondo}. To model analogous effects in the QD region of the three-terminal SC transistor, we generalize the ZBW Hamiltonian of Sec.~\ref{subsec:proximity} by adding to~\eqref{eq:zbw} the term
\begin{equation}
H_{\rm{AB}} = U' (d^\dagger_{\uparrow,\rm{A}}
d^{\hphantom{\dagger}}_{\uparrow,\rm{A}} +
d^\dagger_{\downarrow,\rm{A}}
d^{\hphantom{\dagger}}_{\downarrow,\rm{A}})(d^\dagger_{\uparrow,\rm{B}}
d^{\hphantom{\dagger}}_{\uparrow,\rm{B}} + 
d^\dagger_{\downarrow,\rm{B}}
d^{\hphantom{\dagger}}_{\downarrow,\rm{B}}).
\label{eq:Uinter}
\end{equation}

Building upon the results of Sec.~\ref{subsec:proximity}, we fix $\beta=0.4$, which corresponds to a favorable regime for the output barrier, and assess the loop-coupling asymmetries via $\alpha \in \{ 0.0, 0.2, 0.4 \}$. The resulting current-phase relations are shown in Fig.~\ref{fig:Uinter}, whose layout is analogous to that of Fig.~\ref{fig:zbw}: the level separation decreases from left to right as $\varepsilon_{\rm{B}}/\Delta = 3$ towards $0$, while $U'$ is fixed within each row and increases downwards from $U'/\Delta=0$ in Figs.~\ref{fig:Uinter}(a1)-(a4) to $U'/\Delta=1.5$ in Figs.~\ref{fig:Uinter}(d1)-(d4). The uppermost row thus reproduces the $\beta=0.4$ results of Fig.~\ref{fig:zbw} and serves as the reference.

Reading Fig.~\ref{fig:Uinter} along its columns reveals a systematic trend. In the first column, where the two levels are well separated, level $\rm{B}$ is only marginally occupied making $U'$ negligible for the system. Increasing $U'$ in such a scenario thus enhances the critical current in the open regime only slightly. Also the shifts of the $0$-$\pi$ transition to marginally higher $\varphi_2$ are present. However, the already severely suppressed leakage currents in the closed regime are further diminished. Towards the right-hand columns, where the $U'=0$ results of the uppermost row show a distorted open-regime current-phase relation and a sizable leakage, the effect of $U'$ becomes pronounced. Moderate values restore a current-phase relation resembling that of well-separated levels: the $0$-$\pi$ jump reappears, the critical current in the open regime is recovered, and the closed-regime leakage is suppressed for all values of $\alpha$ considered.

This behavior is readily understood. Since level $\rm{A}$ is kept at half-filling, $\langle n_{\rm{A}} \rangle \approx 1$ and the inter-level repulsion shifts level $\rm{B}$ upwards by approximately $U'$, so that $U'$ effectively restores part of the level separation that Sec.~\ref{subsec:proximity} identified as protective. The progressive shift of the $0$-$\pi$ transition with increasing $U'$, most visible in the right-hand columns, is a direct signature of this mechanism. Within the parameter range examined, the inter-level repulsion therefore acts in favor of the transistor operation, which suggests that it needs neither to be suppressed nor compensated in an experimental realization.

\section{Design proposal in 2DEG platforms \label{sec:fabrication}}

The theoretical requirements identified in the preceding sections translate directly into a concrete device architecture that is realizable with existing nanofabrication techniques. Following various studies of multi-terminal Josephson junctions that utilized the combination of electrostatic gateability of superconductivity in hybrid materials~\cite{Coraiola-2023,Coraiola-2024,Coraiola-2024prx}, we thus consider a planar superconductor/semiconductor (SC/SE) heterostructure in which epitaxial Al is grown in situ on a shallow InAs quantum well hosting a two-dimensional electron gas (2DEG) as shown in Fig.~\ref{fig:sc_trans}. Crucially, the 2DEG carrier density and thus all tunneling barriers are continuously tunable by electrostatic gates, enabling the precise barrier control that the resonance condition \eqref{eq:resonance} demands.

The proposed device geometry, as shown in Fig.~\ref{fig:sc_trans}, consists of an SC layer (typically Al) that is selectively removed to expose three arms of the junction. Electrostatic top gates, separated from the heterostructure surface by a thin oxide dielectric (e. g. $\mathrm{Al_2O_3}$ or $\mathrm{HfO_x}$ deposited by atomic layer deposition), locally deplete the 2DEG beneath each junction to define three independently tunable tunneling barriers and confine the central region into a gate-defined QD. Such a fully top-down approach, illustrated schematically in Fig.~\ref{fig:sc_trans}(b), has been demonstrated for two- and three-terminal planar Josephson junctions on the same Al/InAs platform in Refs.~\cite{Coraiola-2023,Coraiola-2024,Coraiola-2024prx,Graziano-2022}, including independent gate control over each terminal-to-dot barrier down to the single-mode or pinch-off regime. Magnetic flux threading the SC loop is controlled by an on-chip flux line or a small external coil, as routinely done in these geometries~\cite{Coraiola-2023}.

The main requirement for the ideal SC transistor mechanism to hold is the presence of a single dominant level in the central region: the resonance condition $\gamma_1 = \gamma_3$ must hold for the transport-active level, and all other levels must be energetically remote or strongly suppressed at the output barrier. In the Al/InAs platform, this is achieved by increasing the output gate voltage to approach pinch-off of terminal~$2$. This simultaneously lowers $\gamma_2$, which keeps the device inside the region of elevated switching ratio identified in Sec.~\ref{sec:resonance_widths}, and suppresses the leakage contribution of higher-lying levels as demonstrated in Sec.~\ref{sec:zbw}. The loop barriers $\gamma_1$ and $\gamma_3$ are then tuned symmetrically to satisfy \eqref{eq:resonance}, while the total hybridization is kept at intermediate strength $\Gamma \approx \Delta$, where the resonance is surrounded by the broadest region of significant transistor yield.

\section{Conclusions \label{sec:conclusions} }

We have demonstrated the working principle of a three-terminal SC transistor switch with an embedded QD level that operates on purely geometric principles. Ideal switching with a diverging ratio $R = J_\mathrm{on}/J_\mathrm{off} \to \infty$ is achieved by threading half a flux quantum through the control loop while tuning the barriers to $\gamma_3 = (1-\gamma_2)/2 = \gamma_1$, i.~e. so that both arms of the loop couple to the QD with equal tunneling strengths. At this point the two loop phasors cancel, the geometric factor $\chi$ becomes independent of the output phase, and the output supercurrent vanishes identically while the loop sustains a purely circulating supercurrent. The device thus operates as a symmetric SQUID at half-flux quantum. Since the condition involves only the geometric current factor $\chi'_2$, it is independent of the Coulomb interaction $U$, the level position $\varepsilon_{\rm{d}}$, the total hybridization $\Gamma$, and temperature, all of which enter solely through the prefactor $\mathcal{J}_T(\chi)$.

Interactions, however, are significant for practical operations where barrier tuning is not ideal. NRG calculations across the full $\gamma_2$-$\gamma_3$ parameter space show that the resonance line is present for any parameters, but only at intermediate hybridization $\Gamma \approx \Delta$ it becomes embedded in a broad region of elevated $R$. Both weaker and stronger coupling reduce this region to a thin isolated line, so that intermediate $\Gamma$ constitutes the optimal operating regime. The structures surrounding the resonance originate in the singlet-doublet QPTs encoded in $\mathcal{J}_T(\chi)$: elevated ratios require phase diagrams in which singlet islands are well separated by doublet regions, whereas striped phase diagrams yield $R \approx 1$. Finite temperature smooths the underlying discontinuity and suppresses the supercurrent amplitude, but does so comparably in the open and closed regimes, leaving the switching ratio essentially intact up to $T/\Delta = 0.08$. The associated reduction of transmitted supercurrents nevertheless sets a practical lower bound on the operating temperature.

Realistic devices host additional transport-relevant levels in the QD region, and since the gates act on barriers rather than on the actual orbitals, a configuration that symmetrizes the loop couplings of the transport-dominant level generically leaves those of the next level asymmetric. Leakage through the closed device is therefore unavoidable. As expected, leakage is negligible as long as the second level remains well separated in energy as modeled by a ZBW approach. Crucially, raising the output tunneling barrier suppresses it considerably. Benignly, such an operating mode is also beneficial for the purely single-level scenario where lower values of $\gamma_2$ show stable resonance width and improve the accessibility of the SC transistor effect. Inter-level Coulomb repulsion was found to act in the same favorable direction, as it shifts the parasitic level upwards and thereby restores part of the protective level separation.

Taken together, these results define a compact set of design rules for a gate-tunable device manifesting flux-controlled SC switching behavior. First, the two loop barriers shall be matched by gate tuning, while the total hybridization is kept at intermediate strength $\Gamma \approx \Delta$. Second, the output barrier should be raised relative to the tunneling between the loop terminals and the QD region. Both requirements are within reach of existing nanofabrication for shallow Al/InAs 2DEG platforms, making this geometric switching principle a viable route toward SC transistor elements for quantum circuits and cryogenic classical electronics.
	
\begin{acknowledgements}
	The authors would like to thank Tom\'{a}\v{s} Novotn\'{y} for helpful discussions. DN was supported by Grant  No.~23-05263K of the Czech Science Foundation of the Czech Ministry of Education, Youth and Sports. KW was supported by the European Union and the Ministry of Education, Youth and Sports of the Czech Republic (Project: MSCA Fellowships CZ FZU IV - CZ.02.01.01/00/22\_010/0013442). FK acknowledges support by Lumina Quaeruntur fellowship LQ100102602 of the Czech Academy of Sciences. WB acknowledges support by the DAAD PPP (Projekt-ID: 57753337) and the Deutsche Forschungsgemeinschaft (DFG; German Research Foundation) by SFB 1432 (Project No. 425217212). PZ acknowledges the support by the Czech Republic-Germany Mobility programme (ID:8J25DE001) and by the TERAFIT project - CZ.02.01.01/00/22\_008/0004594. 
\end{acknowledgements}

\section*{Data availability}
The data that support the findings of this article are openly available~\cite{2026-transistor_zenodo}.

\bibliography{
	ab.bib,
	abs_ysr.bib,
	add.bib,
	andreev_molecule.bib,
	cmos.bib,
	cps.bib,
    dqd.bib,
	effective.bib,	
	fabrication.bib,	
	frg.bib,	
	gapped_anderson.bib,
	minimal_Kitaev.bib,
	multi_terminal.bib,
	josephson.bib,
	kondo.bib,
    poznan.bib,
	novotny.bib,
	nrg.bib,
	preprints.bib,
	sort.bib,
	sc_surface.bib,
	sc_transistor.bib,
	zalom.bib,
	zenodo.bib,
	zitko.bib
}{}

@article{Herrero-2006Quantum,
  author  = {Jarillo-Herrero, Pablo and van Dam, Jorden A. and Kouwenhoven, Leo P.},
  title   = {Quantum supercurrent transistors in carbon nanotubes},
  journal = {Nature},
  volume  = {439},
  pages   = {953--956},
  year    = {2006},
  doi     = {10.1038/nature04550}
}

@article{vanDam-2006Supercurrent,
  author  = {van Dam, Jorden A. and Nazarov, Yuli V. and Bakkers, Erik P. A. M. and De Franceschi, Silvano and Kouwenhoven, Leo P.},
  title   = {Supercurrent reversal in quantum dots},
  journal = {Nature},
  volume  = {442},
  pages   = {667--670},
  year    = {2006},
  doi     = {10.1038/nature05018}
}

@article{Cleuziou-2006cnt,
  author  = {Cleuziou, J.-P. and Wernsdorfer, W. and Bouchiat, V. and Ondar{\c{c}}uhu, T. and Monthioux, M.},
  title   = {Carbon nanotube superconducting quantum interference device},
  journal = {Nat. Nanotechnol.},
  volume  = {1},
  number  = {1},
  pages   = {53--59},
  year    = {2006},
  doi     = {10.1038/nnano.2006.54}
}

@article{Cleuziou-2007GateTuned,
  author  = {Cleuziou, J.-P. and Wernsdorfer, W. and Andergassen, S. and Florens, S. and Bouchiat, V. and Ondar{\c{c}}uhu, Th. and Monthioux, M.},
  title   = {Gate-Tuned High Frequency Response of Carbon Nanotube {Josephson} Junctions},
  journal = {Phys. Rev. Lett.},
  volume  = {99},
  number  = {11},
  pages   = {117001},
  year    = {2007},
  doi     = {10.1103/PhysRevLett.99.117001}
}

@article{Jorgensen-2006FabryPerot,
  author  = {J{\o}rgensen, H. I. and Grove-Rasmussen, K. and Novotn{\'y}, T. and Flensberg, K. and Lindelof, P. E.},
  title   = {Electron Transport in Single-Wall Carbon Nanotube Weak Links in the {Fabry}--{P}\'{e}rot Regime},
  journal = {Phys. Rev. Lett.},
  volume  = {96},
  number  = {20},
  pages   = {207003},
  year    = {2006},
  doi     = {10.1103/PhysRevLett.96.207003}
}

@article{Jorgensen-2007CriticalCurrent,
  author  = {J{\o}rgensen, H. Ingerslev and Novotn{\'y}, T. and Grove-Rasmussen, K. and Flensberg, K. and Lindelof, P. E.},
  title   = {Critical Current 0--{$\pi$} Transition in Designed {Josephson} Quantum Dot Junctions},
  journal = {Nano Lett.},
  volume  = {7},
  number  = {8},
  pages   = {2441--2445},
  year    = {2007},
  doi     = {10.1021/nl071152w}
}

@article{Ruf-2024GateControl,
  author  = {Ruf, Leon and Puglia, Claudio and Elalaily, Tosson and De Simoni, Giorgio and Joint, Francois and Berke, Martin and Koch, Jennifer and Iorio, Andrea and Khorshidian, Sara and Makk, Peter and Gasparinetti, Simone and Csonka, Szabolcs and Belzig, Wolfgang and Cuoco, Mario and Giazotto, Francesco and Scheer, Elke and Di Bernardo, Angelo},
  title   = {Gate control of superconducting current: {M}echanisms, parameters, and technological potential},
  journal = {Appl. Phys. Rev.},
  volume  = {11},
  number  = {4},
  pages   = {041314},
  year    = {2024},
  doi     = {10.1063/5.0222371}
}

@article{Akazaki-1996JosephsonFET,
  author  = {Akazaki, Tatsushi and Takayanagi, Hideaki and Nitta, Junsaku and Enoki, Takatomo},
  title   = {A {Josephson} field effect transistor using an {InAs}-inserted-channel {In$_{0.52}$Al$_{0.48}$As/In$_{0.53}$Ga$_{0.47}$As} inverted modulation-doped structure},
  journal = {Appl. Phys. Lett.},
  volume  = {68},
  number  = {3},
  pages   = {418--420},
  year    = {1996},
  doi     = {10.1063/1.116704}
}

@article{Pauka-2021,
author={Pauka, S. J. and Das, K. and Kalra, R. and Moini, A. and Yang, Y. and Trainer, M. and Bousquet, A. and Cantaloube, C. and Dick, N. and Gardner, G. C. and Manfra, M. J. and Reilly, D. J.},
title={A cryogenic {CMOS} chip for generating control signals for multiple qubits},
journal={Nat. Electron.},
year={2021},
month={Jan},
day={01},
volume={4},
number={1},
pages={64-70},
issn={2520-1131},
doi={10.1038/s41928-020-00528-y},
url={https://doi.org/10.1038/s41928-020-00528-y}
}

@article{Bohr-2017,
  author={Bohr, Mark T. and Young, Ian A.},
  journal={IEEE Micro}, 
  title={{CMOS} Scaling Trends and Beyond}, 
  year={2017},
  volume={37},
  number={6},
  pages={20-29},
  doi={10.1109/MM.2017.4241347}
  }

@article{
Lundstrom-2003,
author = {Mark Lundstrom },
title = {{Moore}'s Law Forever?},
journal = {Science},
volume = {299},
number = {5604},
pages = {210-211},
year = {2003},
doi = {10.1126/science.1079567}
}

@article{Wrzesniewski2017-kondo,
  title = {{Kondo} physics in double quantum dot based {Cooper} pair splitters},
  author = {Wrze\ifmmode \acute{s}\else \'{s}\fi{}niewski, Kacper and Weymann, Ireneusz},
  journal = {Phys. Rev. B},
  volume = {96},
  issue = {19},
  pages = {195409},
  numpages = {16},
  year = {2017},
  month = {Nov},
  publisher = {American Physical Society},
  doi = {10.1103/PhysRevB.96.195409},
  url = {https://link.aps.org/doi/10.1103/PhysRevB.96.195409}
}

@article{Luethi-2023,
  title = {Planar {Josephson} junctions in germanium: {E}ffect of cubic spin-orbit interaction},
  author = {Luethi, Melina and Laubscher, Katharina and Bosco, Stefano and Loss, Daniel and Klinovaja, Jelena},
  journal = {Phys. Rev. B},
  volume = {107},
  issue = {3},
  pages = {035435},
  numpages = {19},
  year = {2023},
  month = {Jan},
  publisher = {American Physical Society},
  doi = {10.1103/PhysRevB.107.035435},
  url = {https://link.aps.org/doi/10.1103/PhysRevB.107.035435}
}

@Article{Vigneau-2019,
author={Vigneau, Florian and Mizokuchi, Raisei and Zanuz, Dante Colao
and Huang, Xuhai and Tan, Susheng and Maurand, Romain and Frolov, Sergey
and Sammak, Amir and Scappucci, Giordano and Lefloch, Francois and De Franceschi, Silvano},
title={Germanium Quantum-Well {Josephson} Field-Effect Transistors and Interferometers},
journal={Nano Lett.},
year={2019},
month={Feb},
day={13},
publisher={American Chemical Society},
volume={19},
number={2},
pages={1023-1027},
issn={1530-6984},
doi={10.1021/acs.nanolett.8b04275},
url={https://doi.org/10.1021/acs.nanolett.8b04275}
}

@Article{Leblanc-2025,
author={Leblanc, Axel and Tangchingchai, Chotivut and Sadre Momtaz, Zahra and Kiyooka, Elyjah and Hartmann, Jean-Michel and Gustavo, Fr{\'e}d{\'e}ric and Thomassin, Jean-Luc and Brun, Boris and Schmitt, Vivien and Zihlmann, Simon and Maurand, Romain and Dumur, {\'E}tienne and De Franceschi, Silvano and Lefloch, Fran{\c{c}}ois},
title={Gate- and flux-tunable sin(2$\phi$) {Josephson} element with planar-{Ge} junctions},
journal={Nat. Commun.},
year={2025},
month={Jan},
day={25},
volume={16},
number={1},
pages={1010},
issn={2041-1723},
doi={10.1038/s41467-025-56245-7},
url={https://doi.org/10.1038/s41467-025-56245-7}
}

@article{Dolan-1977,
  author  = {Dolan, G. J.},
  title   = {Offset masks for lift-off photoprocessing},
  journal = {Appl. Phys. Lett.},
  volume  = {31},
  pages   = {337--339},
  year    = {1977},
  doi     = {10.1063/1.89690}
}

@article{Likharev-1979,
  author  = {Likharev, K. K.},
  title   = {Superconducting weak links},
  journal = {Rev. Mod. Phys.},
  volume  = {51},
  pages   = {101--159},
  year    = {1979},
  doi     = {10.1103/RevModPhys.51.101}
}

@article{Takayanagi-1985,
  title = {Superconducting Proximity Effect in the Native Inversion Layer on {InAs}},
  author = {Takayanagi, Hideaki and Kawakami, Tsuyoshi},
  journal = {Phys. Rev. Lett.},
  volume = {54},
  issue = {22},
  pages = {2449--2452},
  numpages = {0},
  year = {1985},
  month = {Jun},
  publisher = {American Physical Society},
  doi = {10.1103/PhysRevLett.54.2449},
  url = {https://link.aps.org/doi/10.1103/PhysRevLett.54.2449}
}

@article{Nitta-1993,
  title = {Transport properties in an {InAs}-inserted-channel {${\mathrm{In}}_{0.52}$${\mathrm{Al}}_{0.48}$As/${\mathrm{In}}_{0.53}$${\mathrm{Ga}}_{0.47}$As} heterostructure coupled superconducting junction},
  author = {Nitta, Junsaku and Akazaki, Tatsushi and Takayanagi, Hideaki and Arai, Kunihiro},
  journal = {Phys. Rev. B},
  volume = {46},
  issue = {21},
  pages = {14286--14289},
  numpages = {0},
  year = {1992},
  month = {Dec},
  publisher = {American Physical Society},
  doi = {10.1103/PhysRevB.46.14286},
  url = {https://link.aps.org/doi/10.1103/PhysRevB.46.14286}
}

@article{Doh-2005,
  author  = {Doh, Y.-J. and van Dam, J. A. and Roest, A. L. and Bakkers, E. P. A. M.
             and Kouwenhoven, L. P. and De Franceschi, S.},
  title   = {Tunable supercurrent through semiconductor nanowires},
  journal = {Science},
  volume  = {309},
  pages   = {272--275},
  year    = {2005},
  doi     = {10.1126/science.1113523}
}

@article{Shabani-2016,
  title = {Two-dimensional epitaxial superconductor--semiconductor heterostructures: {A} platform for topological superconducting networks},
  author = {Shabani, J. and Kjaergaard, M. and Suominen, H. J. and Kim, Younghyun and Nichele, F. and Pakrouski, K. and Stankevic, T. and Lutchyn, R. M. and Krogstrup, P. and Feidenhans'l, R. and Kraemer, S. and Nayak, C. and Troyer, M. and Marcus, C. M. and Palmstr\o{}m, C. J.},
  journal = {Phys. Rev. B},
  volume = {93},
  issue = {15},
  pages = {155402},
  numpages = {6},
  year = {2016},
  month = {Apr},
  publisher = {American Physical Society},
  doi = {10.1103/PhysRevB.93.155402},
  url = {https://link.aps.org/doi/10.1103/PhysRevB.93.155402}
}

@article{Kjaergaard-2016,
author={Kjaergaard, M. and Nichele, F. and Suominen, H. J. and Nowak, M. P. and Wimmer, M. and Akhmerov, A. R. and Folk, J. A. and Flensberg, K. and Shabani, J. and Palmstr{\o}m, C. J. and Marcus, C. M.},
title={Quantized conductance doubling and hard gap in a two-dimensional semiconductor--superconductor heterostructure},
journal={Nat. Commun.},
year={2016},
month={Sep},
day={29},
volume={7},
number={1},
pages={12841},
issn={2041-1723},
doi={10.1038/ncomms12841},
url={https://doi.org/10.1038/ncomms12841}
}

@Article{Ke-2019,
author={Ke, Chung Ting and Moehle, Christian M. and de Vries, Folkert K.
and Thomas, Candice and Metti, Sara and Guinn, Charles R. and Kallaher, Ray and Lodari, Mario and Scappucci, Giordano and Wang, Tiantian and Diaz, Rosa E. and Gardner, Geoffrey C. and Manfra, Michael J. and Goswami, Srijit},
title={Ballistic superconductivity and tunable $\pi$-junctions in {InSb} quantum wells},
journal={Nat. Commun.},
year={2019},
month={Aug},
day={21},
volume={10},
number={1},
pages={3764},
issn={2041-1723},
doi={10.1038/s41467-019-11742-4},
url={https://doi.org/10.1038/s41467-019-11742-4}
}

@article{Bocquillon-2017,
author={Bocquillon, Erwann and Deacon, Russell S. and Wiedenmann, Jonas
and Leubner, Philipp and Klapwijk, Teunis M. and Br{\"u}ne, Christoph and Ishibashi, Koji and Buhmann, Hartmut and Molenkamp, Laurens W.},
title={Gapless {Andreev} bound states in the quantum spin {Hall} insulator {HgTe}},
journal={Nat. Nanotechnol.},
year={2017},
month={Feb},
day={01},
volume={12},
number={2},
pages={137-143},
issn={1748-3395},
doi={10.1038/nnano.2016.159},
url={https://doi.org/10.1038/nnano.2016.159}
}

@article{Jaklevic-1964,
  title = {Quantum Interference Effects in {Josephson} Tunneling},
  author = {Jaklevic, R. C. and Lambe, John and Silver, A. H. and Mercereau, J. E.},
  journal = {Phys. Rev. Lett.},
  volume = {12},
  issue = {7},
  pages = {159--160},
  numpages = {0},
  year = {1964},
  month = {Feb},
  publisher = {American Physical Society},
  doi = {10.1103/PhysRevLett.12.159},
  url = {https://link.aps.org/doi/10.1103/PhysRevLett.12.159}
}

@article{Zimmerman-1966,
  title = {Macroscopic Quantum Interference Effects through Superconducting Point Contacts},
  author = {Zimmerman, J. E. and Silver, A. H.},
  journal = {Phys. Rev.},
  volume = {141},
  issue = {1},
  pages = {367--375},
  numpages = {0},
  year = {1966},
  month = {Jan},
  publisher = {American Physical Society},
  doi = {10.1103/PhysRev.141.367},
  url = {https://link.aps.org/doi/10.1103/PhysRev.141.367}
}

@article{Fulton-1972,
  title = {Quantum Interference Properties of Double {Josephson} Junctions},
  author = {Fulton, T. A. and Dunkleberger, L. N. and Dynes, R. C.},
  journal = {Phys. Rev. B},
  volume = {6},
  issue = {3},
  pages = {855--875},
  numpages = {0},
  year = {1972},
  month = {Aug},
  publisher = {American Physical Society},
  doi = {10.1103/PhysRevB.6.855},
  url = {https://link.aps.org/doi/10.1103/PhysRevB.6.855}
}

@Article{Tesche-1977,
author={Tesche, Claudia D. and Clarke, John},
title={dc {SQUID}: {N}oise and optimization},
journal={J. Low Temp. Phys.},
year={1977},
month={Nov},
day={01},
volume={29},
number={3},
pages={301-331},
issn={1573-7357},
doi={10.1007/BF00655097},
url={https://doi.org/10.1007/BF00655097}
}

@article{Koch-2007,
  title = {Charge-insensitive qubit design derived from the {Cooper} pair box},
  author = {Koch, Jens and Yu, Terri M. and Gambetta, Jay and Houck, A. A. and Schuster, D. I. and Majer, J. and Blais, Alexandre and Devoret, M. H. and Girvin, S. M. and Schoelkopf, R. J.},
  journal = {Phys. Rev. A},
  volume = {76},
  issue = {4},
  pages = {042319},
  numpages = {19},
  year = {2007},
  month = {Oct},
  publisher = {American Physical Society},
  doi = {10.1103/PhysRevA.76.042319},
  url = {https://link.aps.org/doi/10.1103/PhysRevA.76.042319}
}

@article{Siano-2004,
  title = {{Josephson} Current through a Nanoscale Magnetic Quantum Dot},
  author = {Siano, F. and Egger, R.},
  journal = {Phys. Rev. Lett.},
  volume = {93},
  issue = {4},
  pages = {047002},
  numpages = {4},
  year = {2004},
  month = {Jul},
  publisher = {American Physical Society},
  doi = {10.1103/PhysRevLett.93.047002},
  url = {https://link.aps.org/doi/10.1103/PhysRevLett.93.047002}
}

@article{Luitz-2010,
  title = {Weak-coupling continuous-time quantum {Monte} {Carlo} study of the single impurity and periodic {Anderson} models with $s$-wave superconducting baths},
  author = {Luitz, David J. and Assaad, Fakher F.},
  journal = {Phys. Rev. B},
  volume = {81},
  issue = {2},
  pages = {024509},
  numpages = {16},
  year = {2010},
  month = {Jan},
  publisher = {American Physical Society},
  doi = {10.1103/PhysRevB.81.024509},
  url = {https://link.aps.org/doi/10.1103/PhysRevB.81.024509}
}

@article{Luitz-2012,
	Author = {Luitz, D. J. and Assaad, F. F. and Novotn{\'y}, T. and Karrasch, C. and Meden, V.},
	Date = {2012/05/30/},
	Day = {30},
	Id = {10.1103/PhysRevLett.108.227001},
	J1 = {PRL},
	Journal = {Phys. Rev. Lett.},
	Journal1 = {Phys. Rev. Lett.},
	Month = {05},
	Number = {22},
	Pages = {227001},
	Publisher = {American Physical Society},
	Title = {Understanding the {Josephson} Current through a {Kondo}-Correlated Quantum Dot},
	Ty = {JOUR},
	Url = {http://link.aps.org/doi/10.1103/PhysRevLett.108.227001},
	Volume = {108},
	Year = {2012}}

@Article{Graziano-2022,
author={Graziano, Gino V. and Gupta, Mohit and Pendharkar, Mihir
and Dong, Jason T. and Dempsey, Connor P. and Palmstr{\o}m, Chris
and Pribiag, Vlad S.}, 
title={Selective control of conductance modes in multi-terminal {Josephson} junctions},
journal={Nat. Commun.},
year={2022},
month={Oct},
day={08},
volume={13},
number={1},
pages={5933},
issn={2041-1723},
doi={10.1038/s41467-022-33682-2},
url={https://doi.org/10.1038/s41467-022-33682-2}
}

@article{Melin-2023,
  title = {Quantum interferometer for quartets in superconducting three-terminal {Josephson} junctions},
  author = {M\'elin, R\'egis and Feinberg, Denis},
  journal = {Phys. Rev. B},
  volume = {107},
  issue = {16},
  pages = {L161405},
  numpages = {6},
  year = {2023},
  month = {Apr},
  publisher = {American Physical Society},
  doi = {10.1103/PhysRevB.107.L161405},
  url = {https://link.aps.org/doi/10.1103/PhysRevB.107.L161405}
}

@article{Jonckheere-2013,
  title = {Multipair dc {Josephson} resonances in a biased all-superconducting bijunction},
  author = {Jonckheere, T. and Rech, J. and Martin, T. and Dou\ifmmode \mbox{\c{c}}\else \c{c}\fi{}ot, B. and Feinberg, D. and M\'elin, R.},
  journal = {Phys. Rev. B},
  volume = {87},
  issue = {21},
  pages = {214501},
  numpages = {6},
  year = {2013},
  month = {Jun},
  publisher = {American Physical Society},
  doi = {10.1103/PhysRevB.87.214501},
  url = {https://link.aps.org/doi/10.1103/PhysRevB.87.214501}
}

@article{Zhang-2023,
  title = {{Andreev} processes in mesoscopic multiterminal graphene {Josephson} junctions},
  author = {Zhang, Fan and Rashid, Asmaul Smitha and Ahari, Mostafa Tanhayi and Zhang, Wei and Ananthanarayanan, Krishnan Mekkanamkulam and Xiao, Run and de Coster, George J. and Gilbert, Matthew J. and Samarth, Nitin and Kayyalha, Morteza},
  journal = {Phys. Rev. B},
  volume = {107},
  issue = {14},
  pages = {L140503},
  numpages = {6},
  year = {2023},
  month = {Apr},
  publisher = {American Physical Society},
  doi = {10.1103/PhysRevB.107.L140503},
  url = {https://link.aps.org/doi/10.1103/PhysRevB.107.L140503}
}

@Article{Riwar-2016,
author={Riwar, Roman-Pascal
and Houzet, Manuel
and Meyer, Julia S.
and Nazarov, Yuli V.},
title={Multi-terminal {Josephson} junctions as topological matter},
journal={Nat. Commun.},
year={2016},
month={Apr},
day={04},
volume={7},
number={1},
pages={11167},
issn={2041-1723},
doi={10.1038/ncomms11167},
url={https://doi.org/10.1038/ncomms11167}
}

@article{Eriksson-2017,
  title = {Topological transconductance quantization in a four-terminal {Josephson} junction},
  author = {Eriksson, Erik and Riwar, Roman-Pascal and Houzet, Manuel and Meyer, Julia S. and Nazarov, Yuli V.},
  journal = {Phys. Rev. B},
  volume = {95},
  issue = {7},
  pages = {075417},
  numpages = {10},
  year = {2017},
  month = {Feb},
  publisher = {American Physical Society},
  doi = {10.1103/PhysRevB.95.075417},
  url = {https://link.aps.org/doi/10.1103/PhysRevB.95.075417}
}

@article{Pankratova-2020,
  title = {Multiterminal {Josephson} Effect},
  author = {Pankratova, Natalia and Lee, Hanho and Kuzmin, Roman and Wickramasinghe, Kaushini and Mayer, William and Yuan, Joseph and Vavilov, Maxim G. and Shabani, Javad and Manucharyan, Vladimir E.},
  journal = {Phys. Rev. X},
  volume = {10},
  issue = {3},
  pages = {031051},
  numpages = {12},
  year = {2020},
  month = {Sep},
  publisher = {American Physical Society},
  doi = {10.1103/PhysRevX.10.031051},
  url = {https://link.aps.org/doi/10.1103/PhysRevX.10.031051}
}

@Article{Strambini-2016,
author={Strambini, E. and D'Ambrosio, S. and Vischi, F. and Bergeret, F. S. and Nazarov, Yu. V. and Giazotto, F.},
title={The $\omega$-{SQUIPT} as a tool to phase-engineer {Josephson} topological materials},
journal={Nat. Nanotechnol.},
year={2016},
month={Dec},
day={01},
volume={11},
number={12},
pages={1055-1059},
issn={1748-3395},
doi={10.1038/nnano.2016.157},
url={https://doi.org/10.1038/nnano.2016.157}
}

@article{Cohen-2018,
doi = {10.1073/pnas.1800044115},  
url = {https://doi.org/10.1073%2Fpnas.1800044115},  
year = 2018,
month = {jun},  
publisher = {Proceedings of the National Academy of Sciences},  
volume = {115},  
number = {27},  
pages = {6991--6994},
author = {Yonatan Cohen and Yuval Ronen and Jung-Hyun Kang and Moty Heiblum and Denis Feinberg and R{\'{e}}gis M{\'{e}}lin and Hadas Shtrikman},  
title = {Nonlocal supercurrent of quartets in a three-terminal {Josephson} junction},  
journal = {Proc. Natl. Acad. Sci. U.S.A.}
}

@Article{Draelos-2019,
author={Draelos, Anne W.
and Wei, Ming-Tso
and Seredinski, Andrew
and Li, Hengming
and Mehta, Yash
and Watanabe, Kenji
and Taniguchi, Takashi
and Borzenets, Ivan V.
and Amet, Fran{\c{c}}ois
and Finkelstein, Gleb},
title={Supercurrent Flow in Multiterminal Graphene {Josephson} Junctions},
journal={Nano Lett.},
year={2019},
month={Feb},
day={13},
publisher={American Chemical Society},
volume={19},
number={2},
pages={1039-1043},
issn={1530-6984},
doi={10.1021/acs.nanolett.8b04330},
url={https://doi.org/10.1021/acs.nanolett.8b04330}
}

@article{Klees-2020,
  title = {Microwave Spectroscopy Reveals the Quantum Geometric Tensor of Topological {Josephson} Matter},
  author = {Klees, R. L. and Rastelli, G. and Cuevas, J. C. and Belzig, W.},
  journal = {Phys. Rev. Lett.},
  volume = {124},
  issue = {19},
  pages = {197002},
  numpages = {6},
  year = {2020},
  month = {May},
  publisher = {American Physical Society},
  doi = {10.1103/PhysRevLett.124.197002},
  url = {https://link.aps.org/doi/10.1103/PhysRevLett.124.197002}
}

@article{Freyn-2011,
  title = {Production of Nonlocal Quartets and Phase-Sensitive Entanglement in a Superconducting Beam Splitter},
  author = {Freyn, Axel and Dou\c{c}ot, Benoit and Feinberg, Denis and M\'elin, R\'egis},
  journal = {Phys. Rev. Lett.},
  volume = {106},
  issue = {25},
  pages = {257005},
  numpages = {4},
  year = {2011},
  month = {Jun},
  publisher = {American Physical Society},
  doi = {10.1103/PhysRevLett.106.257005},
  url = {https://link.aps.org/doi/10.1103/PhysRevLett.106.257005}
}

@article{Weisbrich-2021,
   title={Second {Chern} Number and Non-{Abelian} {Berry} Phase in Topological Superconducting Systems},
   volume={2},
   ISSN={2691-3399},
   url={http://dx.doi.org/10.1103/PRXQuantum.2.010310},
   DOI={10.1103/prxquantum.2.010310},
   number={1},
   journal={PRX Quantum},
   publisher={American Physical Society (APS)},
   author={Weisbrich, H. and Klees, R. L. and Rastelli, G. and Belzig, W.},
   year={2021},
   month=jan,
   pages = {010310} }

@Article{Coraiola-2023,
author={Coraiola, Marco
and Haxell, Daniel Z.
and Sabonis, Deividas
and Weisbrich, Hannes
and Svetogorov, Aleksandr E.
and Hinderling, Manuel
and {ten Kate}, Sofieke C.
and Cheah, Erik
and Krizek, Filip
and Schott, R{\"u}diger
and Wegscheider, Werner
and Cuevas, Juan Carlos
and Belzig, Wolfgang
and Nichele, Fabrizio},
title={Phase-engineering the {Andreev} band structure of a three-terminal {Josephson} junction},
journal={Nat. Commun.},
year={2023},
month={Oct},
day={25},
volume={14},
number={1},
pages={6784},
issn={2041-1723},
doi={10.1038/s41467-023-42356-6},
url={https://doi.org/10.1038/s41467-023-42356-6}
}

@Article{Coraiola-2024,
author={Coraiola, Marco and Svetogorov, Aleksandr E. and Haxell, Daniel Z. and Sabonis, Deividas and Hinderling, Manuel and {ten Kate}, Sofieke C.
and Cheah, Erik and Krizek, Filip and Schott, R{\"u}diger
and Wegscheider, Werner and Cuevas, Juan Carlos and Belzig, Wolfgang
and Nichele, Fabrizio},
title={Flux-Tunable {Josephson} Diode Effect in a Hybrid Four-Terminal {Josephson} Junction},
journal={ACS Nano},
year={2024},
month={Mar},
day={26},
publisher={American Chemical Society},
volume={18},
number={12},
pages={9221-9231},
issn={1936-0851},
doi={10.1021/acsnano.4c01642},
url={https://doi.org/10.1021/acsnano.4c01642}
}

@article{Coraiola-2024prx,
  title = {Spin-Degeneracy Breaking and Parity Transitions in Three-Terminal {Josephson} Junctions},
  author = {Coraiola, M. and Haxell, D. Z. and Sabonis, D. and Hinderling, M. and {ten Kate}, S. C. and Cheah, E. and Krizek, F. and Schott, R. and Wegscheider, W. and Nichele, F.},
  journal = {Phys. Rev. X},
  volume = {14},
  issue = {3},
  pages = {031024},
  numpages = {9},
  year = {2024},
  month = {Aug},
  publisher = {American Physical Society},
  doi = {10.1103/PhysRevX.14.031024},
  url = {https://link.aps.org/doi/10.1103/PhysRevX.14.031024}
}

@article{Antonelli-2025,
  title = {Exploring the Energy Spectrum of a Four-Terminal {Josephson} Junction: {T}oward Topological {Andreev} Band Structures},
  shorttitle = {Exploring the {{Energy Spectrum}} of a {{Four-Terminal Josephson Junction}}},
  author = {Antonelli, Tommaso and Coraiola, Marco and Ohnmacht, David Christian and Svetogorov, Aleksandr E. and Sabonis, Deividas and {ten Kate}, Sofieke C. and Cheah, Erik and Krizek, Filip and Schott, R{\"u}diger and Cuevas, Juan Carlos and Belzig, Wolfgang and Wegscheider, Werner and Nichele, Fabrizio},
  year = 2025,
  month = sep,
  journal = {Phys. Rev. X},
  volume = {15},
  number = {3},
  pages = {031066},
  publisher = {American Physical Society},
  doi = {10.1103/qd3y-f912},
  urldate = {2026-07-06},
  langid = {english}
}

@article{Satori-1992,
	Author = {Satori,Koji and Shiba,Hiroyuki and Sakai,Osamu and Shimizu,Yukihiro},
	Doi = {10.1143/JPSJ.61.3239},
	Journal = {J. Phys. Soc. Jpn.},
	Number = {9},
	Pages = {3239-3254},
	Title = {Numerical Renormalization Group Study of Magnetic Impurities in Superconductors},
	Url = {https://doi.org/10.1143/JPSJ.61.3239},
	Volume = {61},
	Year = {1992}
	}

@article{Yoshioka-2000,
	Author = {Yoshioka, Tomoki and Ohashi, Yoji},
	Da = {2000/06/15},
	Date = {2000/06/15},
	Doi = {10.1143/JPSJ.69.1812},
	Isbn = {0031-9015},
	Journal = {J. Phys. Soc. Jpn.},
	Journal1 = {J. Phys. Soc. Jpn.},
	M3 = {doi: 10.1143/JPSJ.69.1812},
	Month = {2014/08/25},
	Number = {6},
	Pages = {1812--1823},
	Publisher = {The Physical Society of Japan},
	Title = {Numerical Renormalization Group Studies on Single Impurity {Anderson} Model in Superconductivity: {A} Unified Treatment of Magnetic, Nonmagnetic Impurities, and Resonance Scattering},
	Ty = {JOUR},
	Url = {http://dx.doi.org/10.1143/JPSJ.69.1812},
	Volume = {69},
	Year = {2000},
	Year1 = {2000}}

@misc{Zalom-2026-arx,
      title={Symmetry-Induced {Weyl} Nodes in Interacting Multi-Terminal {Josephson} Junctions}, 
      author={Peter Zalom and Gustavo Diniz and David Christian Ohnmacht and Wolfgang Belzig},
      year={2026},
      eprint={2608.21200},
      archivePrefix={arXiv},
      primaryClass={cond-mat.mes-hall},
      url={https://arxiv.org/abs/2608.21200}, 
}

@article{deSimoni-2018,
  title={Metallic supercurrent field-effect transistor},
  author={De Simoni, Giorgio and Paolucci, Federico and Solinas, Paolo and Strambini, Elia and Giazotto, Francesco},
  journal={Nat. Nanotechnol.},
  volume={13},
  number={9},
  pages={802--805},
  year={2018},
  publisher={Nature Publishing Group UK London}
}

@article{Paolucci-2018,
  title={Ultra-efficient superconducting {Dayem} bridge field-effect transistor},
  author={Paolucci, Federico and De Simoni, Giorgio and Strambini, Elia and Solinas, Paolo and Giazotto, Francesco},
  journal={Nano Lett.},
  volume={18},
  number={7},
  pages={4195--4199},
  year={2018},
  publisher={ACS Publications}
}

@article{deSimoni-2019,
  title={{Josephson} field-effect transistors based on all-metallic {Al/Cu/Al} proximity nanojunctions},
  author={De Simoni, Giorgio and Paolucci, Federico and Puglia, Claudio and Giazotto, Francesco},
  journal={ACS Nano},
  volume={13},
  number={7},
  pages={7871--7876},
  year={2019},
  publisher={ACS Publications}
}

@article{deSimoni-2020,
    author = {De Simoni, G. and Puglia, C. and Giazotto, F.},
    title = {Niobium {Dayem} nano-bridge {Josephson} gate-controlled transistors},
    journal = {Appl. Phys. Lett.},
    volume = {116},
    number = {24},
    pages = {242601},
    year = {2020},
    month = {06},
    issn = {0003-6951},
    doi = {10.1063/5.0011304}
}

@article{Puglia-2020,
  title = {Electrostatic Control of Phase Slips in $\mathrm{Ti}$ {Josephson} Nanotransistors},
  author = {Puglia, C. and De Simoni, G. and Giazotto, F.},
  journal = {Phys. Rev. Appl.},
  volume = {13},
  issue = {5},
  pages = {054026},
  numpages = {11},
  year = {2020},
  month = {May},
  publisher = {American Physical Society},
  doi = {10.1103/PhysRevApplied.13.054026},
  url = {https://link.aps.org/doi/10.1103/PhysRevApplied.13.054026}
}

@Article{Rocci-2020,
author={Rocci, Mirko and De Simoni, Giorgio and Puglia, Claudio and Esposti, Davide Degli and Strambini, Elia and Zannier, Valentina and Sorba, Lucia and Giazotto, Francesco},
title={Gate-Controlled Suspended Titanium Nanobridge Supercurrent Transistor},
journal={ACS Nano},
year={2020},
month={Oct},
day={27},
publisher={American Chemical Society},
volume={14},
number={10},
pages={12621-12628},
issn={1936-0851},
doi={10.1021/acsnano.0c05355},
url={https://doi.org/10.1021/acsnano.0c05355}
}

@Article{Ritter-2021,
author={Ritter, M. F.
and Fuhrer, A.
and Haxell, D. Z.
and Hart, S.
and Gumann, P.
and Riel, H.
and Nichele, F.},
title={A superconducting switch actuated by injection of high-energy electrons},
journal={Nat. Commun.},
year={2021},
month={Feb},
day={24},
volume={12},
number={1},
pages={1266},
doi={10.1038/s41467-021-21231-2},
url={https://doi.org/10.1038/s41467-021-21231-2}
}

@Article{Rocci-2021,
author={Rocci, Mirko
and Suri, Dhavala
and Kamra, Akashdeep
and Vilela, Gilv{\^a}nia
and Takamura, Yota
and Nemes, Norbert M.
and Martinez, Jose L.
and Hernandez, Mar Garcia
and Moodera, Jagadeesh S.},
title={Large Enhancement of Critical Current in Superconducting Devices by Gate Voltage},
journal={Nano Lett.},
year={2021},
month={Jan},
day={13},
publisher={American Chemical Society},
volume={21},
number={1},
pages={216-221},
issn={1530-6984},
doi={10.1021/acs.nanolett.0c03547},
url={https://doi.org/10.1021/acs.nanolett.0c03547}
}

@article{Chirolli-transistor-2021,
  title = {Impact of electrostatic fields in layered crystalline {BCS} superconductors},
  author = {Chirolli, Luca and Cea, Tommaso and Giazotto, Francesco},
  journal = {Phys. Rev. Res.},
  volume = {3},
  issue = {2},
  pages = {023135},
  numpages = {8},
  year = {2021},
  month = {May},
  publisher = {American Physical Society},
  doi = {10.1103/PhysRevResearch.3.023135},
  url = {https://link.aps.org/doi/10.1103/PhysRevResearch.3.023135}
}

@article{Amoretti-2022,
  title = {Destroying superconductivity in thin films with an electric field},
  author = {Amoretti, Andrea and Brattan, Daniel K. and Magnoli, Nicodemo and Martinoia, Luca and Matthaiakakis, Ioannis and Solinas, Paolo},
  journal = {Phys. Rev. Res.},
  volume = {4},
  issue = {3},
  pages = {033211},
  numpages = {9},
  year = {2022},
  month = {Sep},
  publisher = {American Physical Society},
  doi = {10.1103/PhysRevResearch.4.033211},
  url = {https://link.aps.org/doi/10.1103/PhysRevResearch.4.033211}
}

@article{Zaccone-2024,
  title = {Theory of superconductivity in thin films under an external electric field},
  author = {Zaccone, Alessio and Fomin, Vladimir M.},
  journal = {Phys. Rev. B},
  volume = {109},
  issue = {14},
  pages = {144520},
  numpages = {7},
  year = {2024},
  month = {Apr},
  publisher = {American Physical Society},
  doi = {10.1103/PhysRevB.109.144520},
  url = {https://link.aps.org/doi/10.1103/PhysRevB.109.144520}
}

@article{Chakraborty-2023,
  title = {Microscopic theory of supercurrent suppression by gate-controlled surface depairing},
  author = {Chakraborty, Subrata and Nikoli\'{c}, Danilo and Cuevas, Juan Carlos and Giazotto, Francesco and Di Bernardo, Angelo and Scheer, Elke and Cuoco, Mario and Belzig, Wolfgang},
  journal = {Phys. Rev. B},
  volume = {108},
  issue = {18},
  pages = {184508},
  numpages = {8},
  year = {2023},
  month = {Nov},
  publisher = {American Physical Society},
  doi = {10.1103/PhysRevB.108.184508},
  url = {https://link.aps.org/doi/10.1103/PhysRevB.108.184508}
}

@article{Bours-2020,
  title = {Unveiling mechanisms of electric field effects on superconductors by a magnetic field response},
  author = {Bours, Lennart and Mercaldo, Maria Teresa and Cuoco, Mario and Strambini, Elia and Giazotto, Francesco},
  journal = {Phys. Rev. Res.},
  volume = {2},
  issue = {3},
  pages = {033353},
  numpages = {11},
  year = {2020},
  month = {Sep},
  publisher = {American Physical Society},
  doi = {10.1103/PhysRevResearch.2.033353},
  url = {https://link.aps.org/doi/10.1103/PhysRevResearch.2.033353}
}

@article{Logan-2014,
	Author = {Logan, David E. and Tucker, Adam P. and Galpin, Martin R.},
	Date = {2014/08/28/},
	Day = {28},
	Id = {10.1103/PhysRevB.90.075150},
	J1 = {PRB},
	Journal = {Phys. Rev. B},
	Journal1 = {Phys. Rev. B},
	Month = {08},
	Number = {7},
	Pages = {075150},
	Publisher = {American Physical Society},
	Title = {Common non-{Fermi} liquid phases in quantum impurity physics},
	Ty = {JOUR},
	Url = {http://link.aps.org/doi/10.1103/PhysRevB.90.075150},
	Volume = {90},
	Year = {2014}
}

@article{Kasumov-1999,
	Author = {Kasumov, A. Yu. and Deblock, R. and Kociak, M. and Reulet, B. and Bouchiat, H. and Khodos, I. I. and Gorbatov, Yu. B. and Volkov, V. T. and Journet, C. and Burghard, M.},
	Date = {1999/05/28},
	Journal = {Science},
	Journal1 = {Science},
	Month = {05},
	Number = {5419},
	Pages = {1508--1511},
	Title = {Supercurrents Through Single-Walled Carbon Nanotubes},
	Ty = {JOUR},
	Url = {http://www.sciencemag.org/content/284/5419/1508},
	Volume = {284},
	Year = {1999}
}

@article{Zalom-2023,
  title = {Rigorous {Wilsonian} renormalization group for impurity models with a spectral gap},
  author = {Zalom, Peter},
  journal = {Phys. Rev. B},
  volume = {108},
  issue = {19},
  pages = {195123},
  numpages = {17},
  year = {2023},
  month = {Nov},
  publisher = {American Physical Society},
  doi = {10.1103/PhysRevB.108.195123},
  url = {https://link.aps.org/doi/10.1103/PhysRevB.108.195123}
}

@article{Zalom-2024,
  title = {Hidden Symmetry in Interacting-Quantum-Dot-Based Multiterminal {Josephson} Junctions},
  author = {Zalom, Peter and \v{Z}onda, M. and Novotn\'y, T.},
  journal = {Phys. Rev. Lett.},
  volume = {132},
  issue = {12},
  pages = {126505},
  numpages = {6},
  year = {2024},
  month = {Mar},
  publisher = {American Physical Society},
  doi = {10.1103/PhysRevLett.132.126505},
  url = {https://link.aps.org/doi/10.1103/PhysRevLett.132.126505}
}

@article{Zalom-2026,
  title = {{Andreev} bound state spectroscopy of a quantum-dot-based {Aharonov--Bohm} interferometer with superconducting terminals},
  author = {Zalom, Peter and Rolih, Don and \ifmmode \check{Z}\else \v{Z}\fi{}itko, Rok},
  journal = {Phys. Rev. B},
  volume = {113},
  issue = {7},
  pages = {075130},
  numpages = {16},
  year = {2026},
  month = {Feb},
  publisher = {American Physical Society},
  doi = {10.1103/32tz-hmwr},
  url = {https://link.aps.org/doi/10.1103/32tz-hmwr}
}

@unpublished{2026-transistor_zenodo,
  author = {Nov\'{a}k, David and Wrze\'{s}niewski, Kacper and Krizek, Filip and Belzig, Wolfgang and Zalom, Peter},
  title = {Data for the manuscript ``{Gate}-tunable supercurrent switching in a three-terminal quantum dot transistor''},
  year = 2026,
  publisher = {Zenodo},
  doi = {10.5281/zenodo.23983552},
  url = {https://doi.org/10.5281/zenodo.23983552},
  note = {\url{https://doi.org/10.5281/zenodo.23983552}}
}

@unpublished{NRGzenodo,
  doi = {10.5281/zenodo.4841076},
  url = {https://doi.org/10.5281/zenodo.4841076},
  author = {\v{Z}itko,  Rok},
  title = {{NRG Ljubljana} (8f90ac4)},
  publisher = {Zenodo},
  year = {2021},
  
  copyright = {GNU Affero General Public License v3.0 or later},
  note = {\url{https://doi.org/10.5281/zenodo.4841076}}
}

\end{document}